\documentclass[usegraphicx,usedcolumn,usenatbib]{mn2e}
\usepackage{graphicx}
\usepackage{txfonts}
\usepackage{subfigure}
\usepackage{rotating}
\usepackage{accents}

\newcommand{\aap}{A\&A}

\bibpunct[,]{(}{)}{;}{a}{,}{,}

\begin{document}

\title[Chemical and photometric models for late type galaxies]
{Chemical and photometric evolution models for disk, irregular and low mass galaxies}

\author[Moll\'{a}]
{Mercedes ~Moll\'{a} $^{1,2}$ \thanks{E-mail:mercedes.molla@ciemat.es}\\
 $^{1}$ Departamento de Investigaci\'{o}n B\'{a}sica, CIEMAT, Avda. Complutense 40, E-28040 Madrid. Spain.\\
 $^{2}$ IAG, University of S\~ao Paulo, 05508-900, S\~ao Paulo-SP. Brasil\\
}
\date{Accepted Received ; in original form }

\pagerange{\pageref{firstpage}--\pageref{lastpage}} \pubyear{2011}

\maketitle \label{firstpage}

\begin{abstract}
We summarize the updated set of multiphase chemical evolution models
performed with 44 theoretical radial mass initial distributions and 10
possible values of efficiencies to form molecular clouds and stars. We
present the results about the infall rate histories, the formation of
the disk, and the evolution of the radial distributions of diffuse and
molecular gas surface density, stellar profile, star formation rate
surface density and elemental abundances of C,N, O and Fe, finding
that the radial gradients for these elements begin very steeper, and
flatten with increasing time or decreasing redshift, although the
outer disks always show a certain flattening for all times.  With the
resulting star formation and enrichment histories, we calculate the
spectral energy distributions (SEDs) for each radial region by using
the ones for single stellar populations resulting from the evolutive
synthesis model {\sc popstar}. With these SEDs we may compute finally
the broad band magnitudes and colors radial distributions in the
Johnson and in the SLOAN/SDSS systems which are the main result of
this work. We present the evolution of these brightness and color
profiles with the redshift.

\end{abstract}


\section{Introduction}

Chemical evolution models to study the evolution of spiral galaxies
has been the subject of a high number of works for the last
decades. From the works by \cite{lyn75,tin80,cla87,cla88,som89}, many
other models have been developed to analyze the evolution of a disk
galaxy. The first attempts were performed to interpret the G-dwarf
metallicity distribution and the radial gradient of abundances
\citep{pei79,sha83,fich91,fitz92,vil96,sma97,aff97,este99,este99b,este99c,sma01}
observed in our Milky Way Galaxy (MWG).  A radial decrease of
abundances was soon observed in most of external spiral galaxies, too
\citep[see][ and references therein]{hen99}, although the shape of the
radial distribution changes from galaxy to galaxy, at least when it is
measured in $\rm dex\,kpc^{-1}$ \footnote{Recent results seem
indicate that the radial gradient may be universal for all galaxies
when is measured in ${\rm dex}\,reff^{-1}$ \citep{san13}, $reff$ being
the radius enclosing the half of the total luminosity of a disk
galaxy, or with any other normalization radius, something already
suggested some years ago by \cite{gar98} although the statistical was
not large enough to reach accurate conclusions.}.

It was early evident that it was impossible to reproduce these
observations by using the classical closed box model \citep{pag89}
which relates the metallicity $Z$ of a region with its fraction of gas
over the total mass, (stars, $s$, plus gas, $g$),
$\mu_{g}=g/(g+s)$. Therefore infall or outflows of gas in MWG and
other nearby spirals were soon considered necessary to fit the data.
In fact, such as it was established theoretically by \citet{gotz92}
and \citet{kop94} a radial gradient of abundances may be created only
by 4 possible ways: 1) A radial variation of the Initial Mass Function
(IMF); 2) A variation of the stellar yields along the galactocentric
radius; 3) A star formation rate (SFR) changing with the radius; 4) A
gas infall rate variable, $f$, with radius. The first possibility is
not usually considered as probable, while the second one is already
included in modern models, since the stellar yields are in fact
dependent on metallicity.  Thus, from the seminal works from
\citet{lac85,gus83} and \citet{cla87} most of the models in the
literature \citep{diaz84,mat89,fer92,fer94,car94,pra95,chia97,boi99},
including the multiphase model used in this work, explain the
existence of this radial gradient by the combined effects of a SFR and
an infall of gas which vary with the galactocentric radius in the
Galaxy.

Most of the chemical evolution models of the literature, included some
of the recently published, are, however, only devoted to the MWG,
totally or only for a region of it, halo or bulge
\citep{cos08,tum10,mar10,cai12,tsu12,mic13} or to any other individual
local galaxy as M~31, M~33 or other local dwarf galaxies
\citep{car02,vaz03,car06,mag07,bar08,mag10,mar10,lan10,her11,kan12,rom13,rob13a,rob13b}.
These works perform the models in a {\sl Tailor-Made Models} way, done
by hand for each galaxy or region. There are not models applicable to
any galaxy, except our grid of models shown in \citet[][hereinafter
MD05]{md05} and these ones from \citet{boi99,boi00}, who presented a
wide set of models with two parameters, the total mass or rotation
velocity and the efficiency to form molecular clouds and stars in MD05
and a angular momentum parameter in the last authors grid.

Besides that, these classical numerical chemical evolution models only
compute the masses in the different phases of a region (gas, stars,
elements...) or the different proportions among them. They do not use
to give the corresponding photometric evolution, preventing the
comparison of chemical information with the corresponding stellar one.
There exist a few consistent models which calculate both things
simultaneously in a consistent way, as those from \citet{vaz03,boi00}
or those from \citet[][ hereafter {\sc
galev}]{uta03,bick04,kot09}. The latter, {\sc galev} evolutionary
synthesis models, describe the evolution of stellar populations
including a simultaneous treatment of the chemical evolution of the
gas and of the spectral evolution of the stellar content. These
authors, however, treat each galaxy as a whole for only some typical
galaxies along the Hubble sequence and does not perform the study of
radial profiles of mass, abundances and light simultaneously. The
series of works by \citet{boi99,boi00,pra00,boi01} seems one
of few that give models allowing an analysis of the chemical and
photometric evolution of disk galaxies.

Given the advances in the instrumentation, it is now possible to study
high redshift galaxies as the local ones with spatial resolution
enough good to obtain radial distributions of abundances and of colors
or magnitudes and thus to perform careful studies of the possible
evolution of the different regions of disk galaxies. For instance to
check the existence of radial gradients at other evolutionary times
different than the present \citep{cres10,yuan11,quey12,jon13} and
their evolution with time or redshift is now possible. It is also
possible to compare these gradients with the radial distributions from
the stellar populations to study possible migration effects.  It is
therefore important to have a grid of consistent
chemo-spectro-photometric models which allows the analysis of both
types of data simultaneously.

The main objective of this work is to give the spectro-photometric
evolution of the theoretical galaxies presented in MD05, for which we
have updated the chemical evolution models.  In that work we presented
a grid of chemical evolution models for 440 theoretical galaxies, with
44 different total mass, as defined by its maximum rotation velocity,
and radial mass distributions, and 10 possible values of efficiencies
to form molecular clouds and stars.  Now we have updated these models
by including a bulge region and by using a different relation
mass--life-meantime for stars now following the Padova stellar tracks.
These models do not consider radial flows, nor stars migration since
no dynamical model is included. The possible outflows by supernova
explosions is not included, too. We check that with the continuous
star formation histories resulting of these models, the  supernova
explosions do not appear in sufficient number as to produce the energy
injection necessary to have outflows of mass.  With these chemical
evolution model results, we calculate the spectro-photometric
evolution by using each time-step of the evolutionary model as a
single stellar populations at which we assign a spectrum taken from
the {\sc popstar} evolutionary synthesis models \citep{mol09}.  Our
purpose in to give as a catalogue the evolution of each radial region
of a disk and this way the radial distributions of elemental
abundances, star formation rate, gas and stars will be available along
with the radial profiles of broad band magnitudes for any time of the
calculated evolution.

The work is divided as follows: we summarize the updated chemical
evolution models in Section 2.  Results related with the surface
densities and abundances are given in Section 3.  We describe our
method to calculate the SEDs of these theoretical galaxies and the
corresponding broad band magnitudes and colors in Section 4. The
corresponding spectro-photometric results are shown in Section 5 where
we give a catalog of the evolution of these magnitudes in the
rest-frame of the galaxies.  Some important predictions arise from
these models which are given in the Conclusions.

\section{The chemical evolution model description}

The model shown here are the ones from MD05 and therefore
a more detailed explanation about the computation is given in that work.
We started with a mass of primordial gas in a spherical region
representing a protogalaxy or {\sl halo}. The initial mass within the
protogalaxy is the dynamical mass calculated by means of the rotation
velocity, $V(Radius)$, through the expression \citep{leq83}:
\begin{equation}
M(Radius)=M_{H,0}=2.32\,10^{5}\,Radius\,V^{2}(Radius)
\label{mtot}
\end{equation}
with $M$ in $\mbox{M}_{\odot}$, $Radius$ in kpc and $V$ in km$\rm \,s^{-1}$.
 We used the Universal Rotation Curve from (URC) from \citet{pss96} to
calculate a set of rotation velocity curves $V(Radius)$ and from these
velocity distributions we obtained the mass radial distributions
$M(Radius)$ (see MD for details and Fig.2 showing these
distributions). It was also possible to use those equations to
obtain the scale length of the disk $R_{D}$, the optical radius,
defined as the one where the surface brightness profile is $\rm
25\,mag\,arcsec^{-2}$, which, if disks follow the Freeman's \citep{free70} law, is
$Ropt=3.2 R_{D}$, and the virial radius, which we take as the
galaxy radius $Rgal$. The total mass of the galaxy $Mgal$ is
taken as the mass enclosed in this radius $Rgal$. The expression
for the URC was given by means of the parameter $\lambda=L/{\rm L_{*}}$, the
ratio between the galaxy luminosity, $L$, and the one of the MWG,
$\rm L_{*}$, in band I. This parameter defines the maximum rotation
velocity, $Vmax$ and the radii described above. Thus, we obtained the values
of the maximum rotation velocities and the corresponding parameters
and mass radial distributions for a set of $\lambda$ values such as it
may be seen in Table~1 from MD05.

To the radial distributions of disks calculated by means of Eq.~\ref{mtot}
described above, we have added a region located at the center ($R=0)$ 
to represent a bulge. The total mass of the bulge is assumed as
a 10\% of the total mass of the disk.  The radius of this bulge is
taken as $R_{D}/5$. Both quantities are estimated from the
correlations found by \citet{marc} among the disk and the bulges data.

\subsection{The infall rate: its dependence on the dynamical mass and on the galactocentric radius}


We assume that the gas falls from the halo to the equatorial plane
forming out the disk in a scenario ELS \citep{ELS}. The time-scale of
this process, or collapse-time scale $\tau_{gal,c}$, characteristic of
every galaxy, changes with its total dynamical mass $Mgal$,
following the expression from \citet{gal84}:
\begin{equation}
 \tau_{gal,c}\propto Mgal^{-1/2}\,T,
\end{equation} 
where $Mgal$ is the total mass of the galaxy,  and $T$ is its age.  
We assume all
galaxies begin to form at the same time and evolves for a time of
$T=13.2$\,Gyr.  We use the value of 13.8\, Gyr, given by the
Planck experiment \citep{ade13} for the age of the Universe and
therefore galaxies start to form at a time $t_{start}=0.6\,\rm Gyr$.

Normalizing to MWG, we obtain:
\begin{equation}
\tau_{gal,c}=\tau_{MWG,c}\sqrt{\frac{M_{MWG}}{Mgal}},
\end{equation}
where $M_{MWG}\sim 1.8\,10^{12} \,{\rm M_{\odot}}$ is the total mass
of MWG and $\tau_{MWG,c}= 4$ \,Gyr (see details in the next paragraph)
is the assumed characteristic collapse-time scale for our Galaxy.

The above expression implies that galaxies more massive than MWG form
in a shorter time-scale, that is more rapidly, than the least massive
galaxies which will need more time to form their disks. This
assumption is in agreement with the observations from
\cite{jim05,heav04,perez13} which find that the most massive
galaxies have their stellar populations in place at very early times
while the less massive ones form most of their stars at $z<1$. This is
also in agreement with self-consistent cosmological simulations which
show that a large proportion of massive objects are formed at early
times (high redshift) while the formation of less massive ones is more
extended in time, thus simulating a modern version of the monolithic
collapse scenario ELS.

The calculated collapse-time scale $\tau_{gal,c}$ is assumed that
corresponds to a radial region located in a characteristic radius,
which is $Rc=Ropt/2 \sim 6.5\,\rm kpc$ for the MWG model which uses
the distribution with $\lambda=1.00$ and number 28, with a maximum
rotation velocity $Vmax=200$\,$\rm km\,s^{-1}$.  The value
$\tau_{MWG,c}=4$ \,Gyr was determined by a detailed study of models
for MWG. We performed a large number of chemical evolution models
changing the inputs free-parameters and comparing the results with
many observational data \citep{fer92,fer94} to estimate the best value
(see section 2.1.2). Similar characteristic radii \footnote{All these
radii and values are related with the stellar light and no with the
mass, but we clear that we do not use them in our models except to
define the characteristic radius $Rc$ for each theoretical mass radial
distribution.  The free parameters are selected for the region defined
by this $Rc$ but taking into account that we normalize the values
after a calibration with the Solar Neighborhood model, a change of
this radius would not modify our model results.}  for our grid of
models were given in Table 1 from MD too, with the characteristic
$\tau_{c}$\footnote{From now we will use the expression $\tau_{c}$ for
$\tau_{gal,c}$ in sake of simplicity.} obtained for each galaxy total
mass $Mgal$.

By taking into account that the collapse time scale depends on the
dynamical mass, and that spiral disks show a clear profile of density
with higher values inside than in the outside regions, we may assume
that the infall rate, and therefore the collapse time scale $\tau_{coll}$, 
has a radial dependence, too. Since the mass density
seems to be an exponential in most of cases, we then assumed a similar
expression:
\begin{equation}
\tau_{coll}(Radius)=\tau_{c} \exp{\frac{(Radius-Rc)}{\lambda_{D}}}
\end{equation}
where $\lambda_{D}$ is the scale-length of the
collapse time-scale, taken as around the half of the scale-length of
surface density brightness distribution, $R_{D}$, that is 
$\lambda_{D}=0.15Ropt\sim 0.5 R_{D}$.

\begin{figure}
\centering
\subfigure{\includegraphics[width=0.35\textwidth,angle=-90]{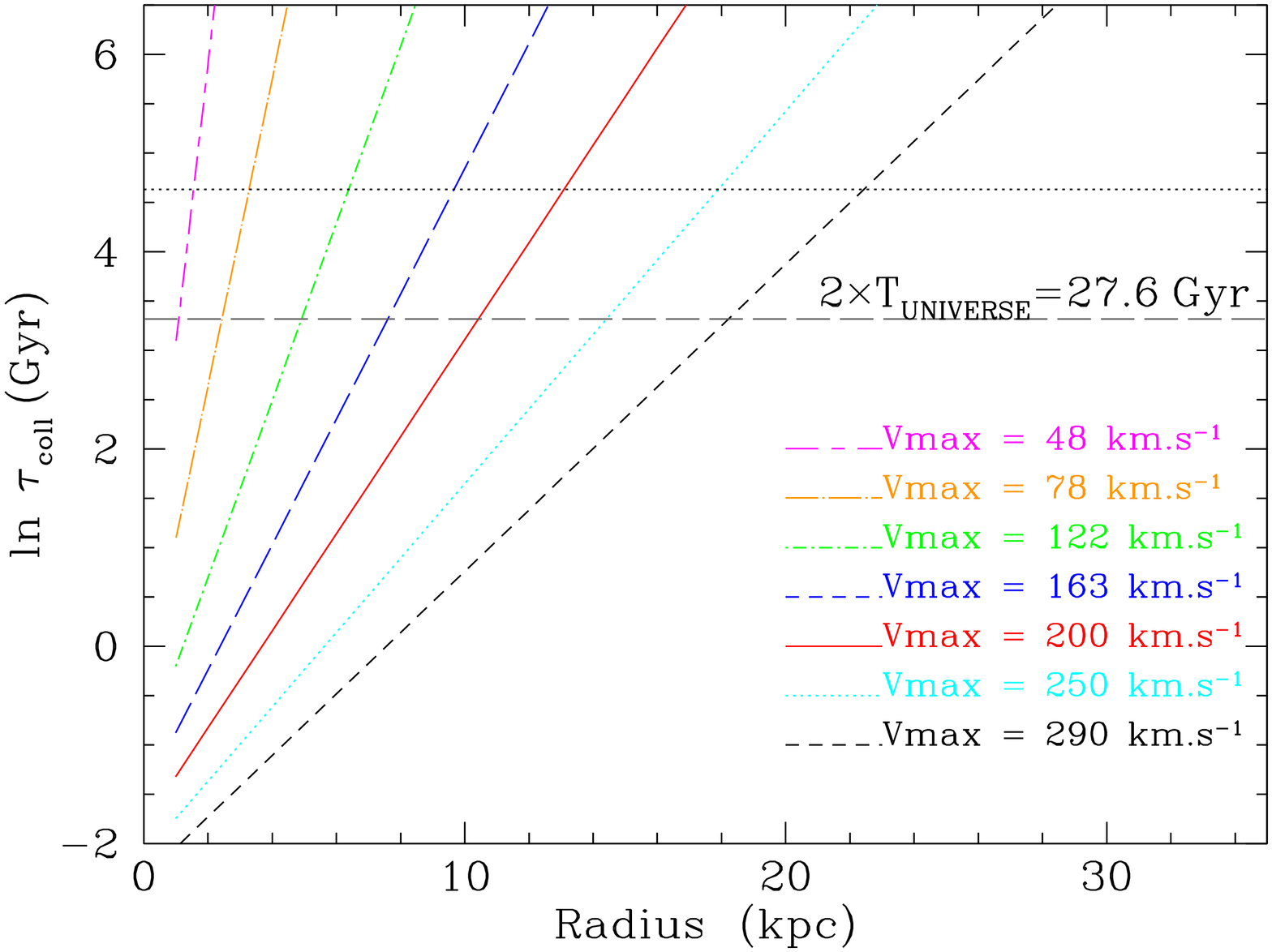}}
\subfigure{\includegraphics[width=0.35\textwidth,angle=-90]{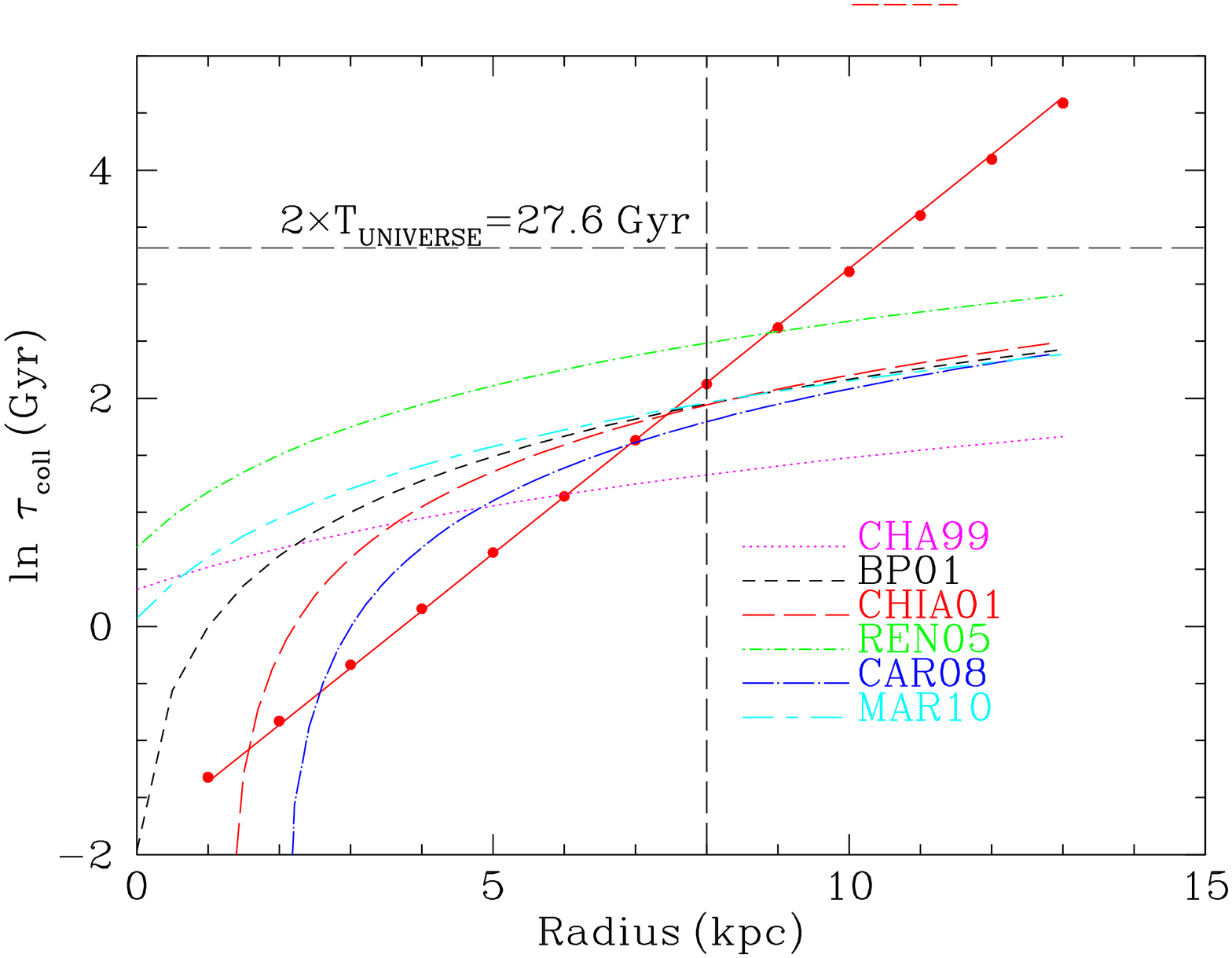}}
\caption{a) Dependence of the collapse time scale, $\tau_{coll}$, in
natural logarithmic scale, (in Gyr) on the galactocentric radius,
$Radius$, in kpc. Each line represents a given maximum rotation
velocity, $Vmax$ or radial mass distribution as labeled. The dashed
(gray) and dotted (black) lines show the time corresponding to 2 and 5
times, respectively, the age of the Universe, $T_{UNIVERSE}= 13.8$\,
Gyr.  b) Comparison of the radial dependence of the collapse time
scale, $\tau_{coll}$, in natural logarithmic scale, for our MWG model,
corresponding to $Vmax=200\,\rm{km\,s^{-1}}$, shown by the solid red
line and labeled MD, with the radial functions used by
\citet{cha99,boi99,chia01,ren05,car08,mar10}, labeled
CHA99,BP99,CHIA01,REN05,CAR08 and MAR10, respectively.}
\label{tcol_r}
\end{figure}

Obviously the collapse time scale for the bulge region is obtained
naturally from the above equation with $R=0$.  We show in the upper
panel of Fig.~\ref{tcol_r} the collapse time scale $\tau_{coll}$, in
natural logarithmic scale, as a function of the galactocentric radius,
for seven radial distributions of total mass, as defined by their
maximum rotation velocity, $Vmax$, and plotted with different color,
as labeled. These seven theoretical galaxies are used as examples and
their characteristics are summarized in Table~\ref{examples}, where we
have the number of the distribution, $\rm dis$, corresponding to
column 2 from Table 1 in MD in column 1, the maximum rotation
velocity, $Vmax$, in $\rm km\,s^{-1}$, in column 2, the total
mass,$Mgal$, in $10^{11}\, \rm M_{\odot}$ units, in column 3, the
theoretical optical radius $Ropt$, following \citet{pss96} equations,
in kpc, in column 4, the collapse time scale in the characteristic
radius, $\tau_{c}$, in Gyr in column 5, the value $nt$ which defines
the efficiencies (see Eq.22 and 23 in section 2.2) in column 6, and
the values for these efficiencies in columns 7 and 8.

The red line corresponds to a MWG-like radial distribution. The
long-dashed black line shows the time corresponding to 2 times the age
of the Universe.  Such as we may see, the most massive galaxies would
have the most extended disks, since the collapse timescale is smaller
than the age of the universe for longer radii, thus allowing the
formation of the disk until radii as larger as 20\,kpc, while the
least massive ones would only have time to form the central region,
smaller than 1-2\,kpc, as observed. The dashed (gray) lines show the
time corresponding to 2 times the age of the Universe, $T_{UNIVERSE}=
13.8$\, Gyr. The dotted black line defines the collapse time scale for
which the maximum radius for the disk of the MWG model would be
13\,kpc, the optical radius and it corresponds to a collapse time
scale of 5 times the age of the Universe.

\begin{table}
\caption{Theoretical galaxy models selected to represent a simulated Hubble sequence}
\begin{tabular}{ccccccccc}
\hline
dis & $Vmax$ & $Mgal$ & $Ropt$  & $\tau_{c}$ & $nt$ & $\epsilon_{\nu}$ & $\epsilon_{\delta}$\\
   &$\rm km\,s^{-1}$ & $\rm 10^{11}\,M_{\odot}$ & kpc & Gyr &  &  &    \\
 3 &  48 &  0.3 &  2.3 &  31.6 & 8 & 0.037 & 2.6 10$^{-4}$  \\
10 &  78 &  1.3 &  4.1 &  15.5 & 7 & 0.075 & 1.5 10$^{-3}$  \\
21 & 122 &  4.3 &  7.1 &   8.1 & 6 & 0.15  & 1.0 10$^{-2}$ \\
24 & 163 &  9.8 & 10.1 &   5.4 & 5 & 0.30  & 5.0 10$^{-2}$ \\
28 & 200 & 17.9 & 13.0 &   4.0 & 4 & 0.45  & 1.4 10$^{-1}$  \\
35 & 250 & 33.5 & 16.9 &   2.9 & 3 & 0.65  & 3.4 10$^{-1}$ \\
39 & 290 & 52.7 & 20.6 &   2.3 & 1 & 0.95  & 8.8 10$^{-1}$ \\ 
\hline  
\label{examples}
\end{tabular}
\end{table}


Other authors have also included a radial dependence for the infall
rate in their models \citep{lac85,mat89,fer92,pcb98,boi00} with
different expressions. In fact this dependence, which produces an
in-out formation of the disk, is essential to obtain the observed
density profiles and the radial gradient of abundances, such as it has
been stated before \citep{mat89,fer94,boi00}. In the bottom panel of
the same Fig.~\ref{tcol_r} we show the collapse time scale
$\tau_{coll}$, in natural logarithmic scale, assumed in different
chemical evolution models of MWG, as a function of the galactocentric
radius. The red solid line corresponds to our MWG model ($\lambda=1.00$
and Number 28 of the mass distributions of Table 1) from MD. The
other functions, straight lines, are those used by
\citet{cha99,boi99,chia01,ren05,car08,mar10}, as labeled.  Since they
use straight lines the collapse time scale for our model results
shorter for the inner disk regions (except for the bulge region $R<
3-4\,\rm kpc)$ and longer for the outer ones, than the ones used by
the other works. This will have consequences in the radial
distributions of stars and elemental abundances as we will see.

\subsection{The star formation law in two-steps: the formation of molecular gas phase} 

The star formation is assumed different in the halo than in the disk.
In the halo the star formation follows a Kennicutt-Schmidt law.  In
the disk, however, we assume a more complicated star formation law, by
creating molecular gas from the diffuse gas in a first step, again by
a Kennicutt-Schmidt law. And then stars from from the cloud-cloud
collisions.  There is a second way to create stars from the
interaction of massive stars with the surrounding molecular clouds.

Therefore the equation system of our model is:

\begin{eqnarray}
\frac{dg_{H}}{dt}&=&-\left(\kappa_{1}+\kappa_{2}\right)g^{n}_{H}-fg_{H}+W_{H}\\
 \frac{ds_{1,H}}{dt}&=&\kappa_{1}g^{n}_{H}-D_{1,H}\\
 \frac{ds_{2,H}}{dt}&=&\kappa_{2}g^{n}_{H}-D_{2,H}\\
 \frac{dg_{D}}{dt}&=&-\eta g^{n}_{D}+\alpha'c_{D}s_{2,D}+\delta'c^{2}_{D}+fg_{H}+W_{D}\\
 \frac{dc_{D}}{dt}&=&\eta g^{n}_{D}-\left(\alpha_{1}+\alpha_{2}+\alpha'\right)c_{D}s_{2D}-\left(\delta_{1}+\delta_{2}+\delta'\right)c^{2}_{D}\\
  \frac{ds_{1,D}}{dt}&=&\delta_{1}c^{2}_{D}+\alpha_{1}c_{D}s_{2D}-D_{1,D}\\
 \frac{ds_{2,D}}{dt}&=&\delta_{2}c^{2}_{D}+\alpha_{2}c_{D}s_{2D}-D_{2,D}\\
 \frac{dr_{H}}{dt}&=&D_{1,H}+D_{2,H}-W_{H}\\
 \frac{dr_{D}}{dt}&=&D_{1,D}+D_{2,D}-W_{D}\\
 \frac{X_{i,H}}{dt}&=&\frac{(W_{i,H}-X_{i,H}W_{H})}{g_{H}}\\
 \frac{X_{i,D}}{dt}&=&\frac{[W_{i,D}-X_{i,D}W_{D}+fg_{H}(X_{i,H}-X_{i,D})]}{g_{D}+c_{D}}
\end{eqnarray}
 
These equations predict the time evolution of the different phases of
the model: diffuse gas, $g$, molecular gas, $c$, low mass stars, $s_{1}$, 
and intermediate mass and massive stars, $s_{2}$, and
stellar remnants, $r$, (where letters $D$ and $H$ correspond to disk and
halo, respectively). Stars are divided in 2 ranges, $ s_{1}$
being the low mass stars, and $ s_{2}$ the intermediate and massive ones,
considering the limit between both ranges stellar a mass $m=4\,\rm M_{\odot}$.
$X_{i}$ are the mass fractions of the 15 elements considered by the
model:$^{1}$H, D, $^{3}$He, $^{4}$He, $^{12}$C, $^{16}$O, $^{14}$N,
$^{13}$C, $^{20}$Ne, $^{24}$Mg, $^{28}$Si, $^{32}$S, $^{40}$Ca,
$^{56}$Fe, and the rich neutron isotopes created from $^{12}$C,
$^{16}$O, $^{14}$N and from $^{13}$C.
 
 Therefore we have different processes defined in the galaxy:
  
\begin{enumerate}
\item Star formation by spontaneous fragmentation of gas in the halo:
$\propto \kappa_{1,2}g_{D}^{n}$, where we use $n = 1.5$
 \item Clouds formation by diffuse gas: $\propto \eta g_{D}^{n}$
with $ n= 1.5$, too
 \item Star formation due to cloud collision: $\propto \delta_{1,2}c_{D}^{2}$
 \item Diffuse gas restitution due to cloud collision: $\propto \delta'c_{D}^{2}$
 \item Induced star formation due to the interaction between clouds
and massive stars: $\propto \alpha_{1,2}c_{D}s_{2,D}$
 \item Diffuse gas restitution due to the induced star formation:
$\alpha'c_{D}s_{2,D}$
 \item Galaxy formation by gas accretion from the halo or protogalaxy: $fg_{H}$
\end{enumerate}

where $\alpha$, $\delta$, $\eta$ and $\kappa$ are the proportionality factors of
the stars and cloud formation and are free input parameters. \footnote{Since stars are divided in two groups: those with
$s_{1}$, and $s_{2}$, the
parameters are divided in the two groups too, thus
$\kappa=\kappa_{1}+\kappa_{2}$, $\delta=\delta_{1}+\delta_{2}$,
$\alpha=\alpha_{1}+\alpha_{2}$. }.

Thus, the star formation law in halo and disk is: 
 \begin{eqnarray}
 \Psi_{H}(t)& =& (\kappa_{1}+\kappa_{2})g_{H}^{n}\\
 \Psi_{D}(t) &=& (\eta_{1}+\eta_{2})c_{D}^{2}+(\alpha_{1}+\alpha_{2})c_{D}s_{2D}
\end{eqnarray}
 
Although the number of parameters seems to be large, actually not all
of them are free. For example, the infall rate, $f$, is the inverse of
the collapse time $\tau_{coll}$, as we described in the above section;
Proportionality factors $\kappa$, $\eta$, $\delta$ and $\alpha$ have
a radial dependence, as we show in the study of MWG \citet{fer94},
which may be used in all disks galaxies through the volume of each radial region
and some proportionality factors called {\sl efficiencies}.
These efficiencies or proportionality factors of these equations have
a probability meaning and therefore their values are in the range
[0,1].  The efficiencies are then: the probability of star formation
in the halo, $\epsilon_{\kappa}$, the probability of cloud formation,
$\epsilon_{\eta}$; cloud collision, $\epsilon_{\delta}$; and the
interaction between massive stars $\epsilon_{\alpha}$. This last one
has a constant value since it corresponds to a local process. The
efficiency to form stars in the halo is also assumed constant for all
of them. Thus, the number of free parameters is reduced to
$\epsilon_{\eta}$ and $\epsilon_{\delta}$.

The efficiency to form stars in the halo, $\epsilon_{\kappa}$, is
obtained through the selection of the best value $\kappa$ to reproduce
the SFR and abundances of the Galactic halo \citep[see][for
details]{fer94}. We assumed that it is approximately constant for all
halos with a value $\epsilon_{\kappa}\sim 1.5-6 \,10^{-3}$. The value
for $\epsilon_{\alpha}$ is also obtained from the best
value $\alpha$ for MWG and assumed as constant for all galaxies since
these interactions massive stars-clouds are local processes. The other
efficiencies $\epsilon_{\eta}$ and $\epsilon_{\delta}$ may take any
value in the range [0-1]. From our previous models calculated for
external galaxies of different types \citep{mol96}, we found that both
efficiencies must change simultaneously in order to reproduce the
observations, with higher values for the earlier morphological types
and smaller for the the later ones. In MD05 there is a clear
description about the selection of values and the relation
$\epsilon_{\eta}-\epsilon_{\delta}$. As a summary, we have calculated
these efficiencies with the expressions:
\begin{eqnarray}
\epsilon_{\eta} &=& \exp{\frac{nt}{20}}\\
\epsilon_{\delta} & = & \exp{\frac{nt}{8}}
\end{eqnarray}
selecting 10 values $nt$ between 1 and 10 (we suggest to select a value
$nt$ similar to the Hubble type index to obtain model results fitting
the observations).  The efficiencies values computed for the grid from
MD05 are shown in Table 2 from that work.

\subsection{Stellar yields, Initial Mass function and Supernova Ia rates}

The selection of the stellar yields and the IMF needs to be done
simultaneously since the integrated stellar yield for any element,
which defines the absolute level of abundances for a given model,
depends on both ingredients.  In this work we used the IMF from
\citet{fer90} with limits $m_{low}=0.15$ and $m_{up}=100\, \rm
M_{\odot}$. The stellar yields are from \citet{woo95} for massive
stars ($m \ge 8\, \rm M_{\odot})$ and from \citet{gav05,gav06} for low
and intermediate mass stars ($0.8 \,{\rm M_{\odot}} < m \le 8 \,\rm
M_{\odot}$). Stars in the range $0.15\,{\rm M_{\odot}} < m < 0.8 \,\rm
M_{\odot}$ have no time to die, so they still live today and do not
eject any element to the interstellar medium.  The mean stellar
lifetimes are taken from the isochrones from the Padova group
\citep{bre94,fag94a,fag94b,gir96}, instead using those from the Geneva group
\cite{sha92}. This change is done for consistency since we
use the Padova isochrones on the {\sc popstar} code that we will use
for the spectro-photometric models. The supernova Ia yields are taken
from \citet{iwa99}. The combination of these stellar yields with this
IMF produces the adequate level of CNO abundances, able to reproduce
most of observational data in the MWG galaxy \citep{gav05,gav06}, in
particular the relative abundances of C/O, N/O, and C/Fe, O/Fe, N/Fe.
The study of other combinations of IMF and stellar yields will be
analyze in \citet{mol14}. The supernova type Ia rates are calculated
by using prescriptions from \citet{ruiz00}.

\section{Results: Evolution of disks with redshift}

The chemical evolution models are given in Tables~\ref{masas} and
\ref{abun}. We show as an example some lines corresponding to the
model $nt=4$ and $dis=28$, the whole set of results will be given in
electronic format as a catalogue.  In Table~\ref{masas} we give the
type of efficiencies $nt$ and the distribution number $dis$ in columns
1 and 2, the time in Gyr in column 3, the corresponding redshift in
column 4, the radius of each disk region in kpc in column 5, and the
star formation rate in the halo and in the disk, in
$\mbox{M}_{\odot}\,yr^{-1}$ , in columns 6 and 7.  In next columns we
have the total mass in the halo and in the disk. In columns 8 to 16 we
show the mass in each phase of the halo, diffuse gas, low-mass stars,
massive stars, and mass in remnants (columns 8 to 11) and of the disk,
diffuse and molecular gas, low-mass stars, massive stars, and mass in
remnants (columns 12 to 16).

\begin{table*}
\begin{center}
\caption{Evolution of different phases along the time/redshift for the
grid of models. We show as example the results for the present time of
a MWG-like model ($nt=4$, $dis=28$). The complete table will be
available in electronic format.}
\begin{tabular}{ccccccccc}
\hline
$nt$  & $dis$ &   $t$ & $z$ &  $Radius$  & $\Psi_{H}$ & $\Psi_{D}$ & $M_{H}$ & $M_{D}$ \\
        &           &  Gyr &      &    kpc  & $\rm M_{\odot}\,yr^{-1}$ & $\rm M_{\odot}\,yr^{-1}$ & $\rm 10^{9}\,M_{\odot}$ & $\rm 10^{9}\,M_{\odot}$  \\
\hline
 4 & 28 & 1.3201e+01 & 0.00&   0 &  1.5431e-05 &  4.9526e-02 &  1.0337e+00 &  8.4818e+00\\
 4 & 28 & 1.3201e+01 & 0.00&   2 &  2.5875e-10 &  2.1748e-02 &  9.7257e-03 &  4.2363e+00\\
 4 & 28 & 1.3201e+01 & 0.00&   4 &  2.9111e-08 &  5.7276e-02 &  6.4913e-02 &  9.7481e+00\\
 4 & 28 & 1.3201e+01 & 0.00&   6 &  2.1827e-04 &  1.8737e-01 &  3.7799e-01 &  1.2042e+01\\
 4 & 28 & 1.3201e+01 & 0.00&   8 &  9.6978e-03 &  4.7546e-01 &  2.8732e+00 &  9.6868e+00\\
 4 & 28 & 1.3201e+01 & 0.00&  10 &  3.4321e-02 &  3.8606e-01 &  6.6828e+00 &  5.0472e+00\\
 4 & 28 & 1.3201e+01 & 0.00&  12 &  4.9261e-02 &  1.8346e-01 &  8.7949e+00 &  2.0751e+00\\
 4 & 28 & 1.3201e+01 & 0.00&  14 &  5.2833e-02 &  7.0495e-02 &  9.5228e+00 &  7.8723e-01\\
 4 & 28 & 1.3201e+01 & 0.00&  16 &  5.2396e-02 &  2.2987e-02 &  9.7360e+00 &  2.9398e-01\\
 4 & 28 & 1.3201e+01 & 0.00&  18 &  5.1446e-02 &  5.7661e-03 &  9.8290e+00 &  1.1004e-01\\
 4 & 28 & 1.3201e+01 & 0.00&  20 &  5.0964e-02 &  9.0888e-04 &  9.9186e+00 &  4.1360e-02\\
 4 & 28 & 1.3201e+01 & 0.00&  22 &  5.1061e-02 &  7.9634e-05 &  1.0014e+01 &  1.5583e-02\\
 4 & 28 & 1.3201e+01 & 0.00&  24 &  5.1747e-02 &  4.7257e-06 &  1.0104e+01 &  5.8713e-03\\
\hline
\label{masas}
\end{tabular}
\end{center}
\end{table*}

\begin{table*}
\addtocounter{table}{-1}
\begin{center}
\caption{Cont. Evolution of the different phases along the time/redshift for the grid of models}
\begin{tabular}{ccccccccc}
\hline
 $g_{H}$ & $s_{1,H}$ & $s_{2,H}$ & $rem_{H}$ & $g_{D}$ & $c_{D}$ & $s_{1,D}$ & $s_{2,D}$ & $rem_{D}$ \\
  $10^{9}\,\mbox{M}_{\odot}$& $10^{9}\,\mbox{M}_{\odot}$& $10^{9}\,\mbox{M}_{\odot}$& $10^{9}\,\mbox{M}_{\odot}$& $10^{9}\,\mbox{M}_{\odot}$& 
  $10^{9}\,\mbox{M}_{\odot}$& $10^{9}\,\mbox{M}_{\odot}$& $10^{9}\,\mbox{M}_{\odot}$& $10^{9}\mbox{M}_{\odot}$\\
\hline
 2.9016e-03 & 8.6815e-01 & 1.2048e-05 & 1.6259e-01   &  2.7416e-02&  1.4705e+00 & 7.1689e+00 & 3.3978e-04 & 1.2371e+00 \\
 1.3879e-05 & 8.1686e-03 & 1.7090e-07 & 1.5430e-03   &  1.4016e-02&  1.4705e+00 & 3.5667e+00 & 1.5015e-04 & 6.2852e-01 \\
 4.0674e-04 & 5.4436e-02 & 4.7128e-07 & 1.0070e-02   &  4.1804e-02&  1.4705e+00 & 8.2575e+00 & 3.9145e-04 & 1.3868e+00 \\
 1.7779e-01 & 1.7061e-01 & 2.0974e-06 & 2.9597e-02   &  1.3955e-01&  1.4705e+00 & 1.0242e+01 & 1.2864e-03 & 1.5234e+00 \\
 2.4434e+00 & 3.7340e-01 & 6.7054e-05 & 5.6303e-02   &  3.5938e-01&  1.4705e+00 & 8.0530e+00 & 3.2457e-03 & 1.0218e+00 \\
 6.0752e+00 & 5.3545e-01 & 2.3428e-04 & 7.1939e-02   &  3.9304e-01&  1.4705e+00 & 3.9522e+00 & 2.6251e-03 & 4.4801e-01 \\
 8.1503e+00 & 5.7161e-01 & 3.3513e-04 & 7.2706e-02   &  2.8478e-01&  1.4705e+00 & 1.4482e+00 & 1.2453e-03 & 1.5079e-01 \\
 8.9024e+00 & 5.5139e-01 & 3.5900e-04 & 6.8586e-02   &  1.7444e-01&  1.4705e+00 & 4.4266e-01 & 4.7796e-04 & 4.2150e-02 \\
 9.1454e+00 & 5.2546e-01 & 3.5588e-04 & 6.4789e-02   &  9.5622e-02&  1.4705e+00 & 1.1078e-01 & 1.5556e-04 & 9.4763e-03 \\
 9.2582e+00 & 5.0801e-01 & 3.4936e-04 & 6.2419e-02   &  4.6570e-02&  1.4705e+00 & 2.0439e-02 & 3.8890e-05 & 1.5411e-03 \\
 9.3567e+00 & 5.0019e-01 & 3.4607e-04 & 6.1374e-02   &  2.1461e-02&  1.4705e+00 & 2.3765e-03 & 6.0946e-06 & 1.5895e-04 \\
 9.4528e+00 & 4.9999e-01 & 3.4672e-04 & 6.1317e-02   &  1.0009e-02&  1.4705e+00 & 1.7210e-04 & 5.3111e-07 & 1.0686e-05 \\
 9.5352e+00 & 5.0643e-01 & 3.5137e-04 & 6.2099e-02   &  4.4897e-03&  1.4705e+00 & 9.3097e-06 & 3.1422e-08 & 5.5691e-07 \\
\hline
\end{tabular}
\end{center}
\end{table*}

We will give the results obtained for our grid of models by showing
the corresponding ones to the galaxies from Table~\ref{examples} as a
function of the redshift or of the galactocentric radius. When the
radial distributions are plotted, we do that for several times or
values of redshifts.  We assume that each time in the evolution of a
galaxy corresponds to a redshift. To calculate this redshift, we use
the relation redshift-evolutionary time given by \cite{mac06} with the
cosmological parameters from the same PLANK experiment \cite{ade13}
($\Omega_\lambda=0.685, H_{0}=67.3$), and this way the time assumed
for the beginning of the galaxy formation, $t_{start}=0.6\,\rm Gyr$ ,
corresponds to a redshift $z=7$.

\subsection{The formation of the disk}

\begin{figure}
\includegraphics[width=0.35\textwidth,angle=-90]{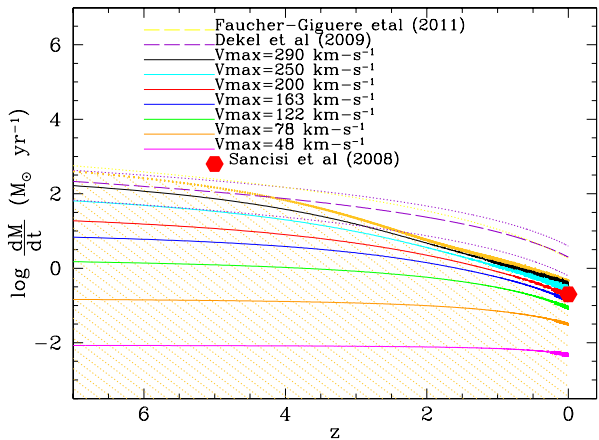}
\caption{The evolution of the infall gas rate along the redshift $z$
for all radial regions of our 44 galaxy mass values represented by the
shaded zone.  The solid lines show the evolution of the whole infall
rate of the same 7 theoretical galaxies shown in Fig.~1 with the same
color coding.  The highest the total mass, the highest the infall
rate. Dashed purple and yellow lines represent the prescriptions from
\citet{dek09} and \citet{fau11}, respectively. The dotted purple lines
are the \citet{dek09} prescriptions for masses $2\,10^{12}\,\rm
M_{\odot}$ and $3\,10^{11}\,\rm M_{\odot}$, while the red hexagon in
the present time is the estimated value given by \citet{san08}.}
\label{infall_z}
\end{figure}

The process of infall of gas from the protogalaxy to the equatorial
plane depends on time since the mass remaining in the protogalaxy or
halo is decreasing with time. In Fig.~\ref{infall_z} we represent the
resulting total infall of mass for our 44 radial distributions of mass
as a function of the redshift $z$. Since this process is defined by
the collapse time-scale, the total infall rate for the whole galaxy
only depends on the total mass, and therefore there would be only 44
possible results, one for each maximum rotation velocity, or mass of
the theoretical galaxy. However, since we have assumed an infall rate
variable with the radius, each radial region of a galaxy has a
different infall rate.  We show as a shaded region the locus where our
results for all radial regions of the whole set of models fall.  Over
this region we show as solid lines the results for the whole infall of
the same 7 theoretical galaxies as in the previous Fig.~\ref{tcol_r},
and with the same color coding, as labeled.  The dashed purple and
yellow lines correspond to the prescriptions given by \citet{dek09}
and \citet{fau11} for the infall of gas as obtained by their
cosmological simulations to form massive galaxies. In both cases these
expressions depend on the dynamical halo mass, so we have shown both
lines for $Mdyn=10^{12} \,\rm M_{\odot}$ which will be compared to our
most massive model which is on the top (black line) and which has with
a similar total mass.  There are some differences with the results
from cosmological simulations for the lowest redshifts, for which our
models have lower infall rates than these cosmological simulations.
However we remind that these simulations prescriptions are valid for
spheroidal galaxies. We show as dotted purple lines 2 other lines
following \citet{dek09} prescriptions but for masses $2\,10^{12}\,\rm
M_{\odot}$ and $3\,10^{11}\,\rm M_{\odot}$, above and below the
standard dashed line, checking that this last one is very similar to
our cyan line. Therefore to decrease the infall rate for the most
recent times is probably a good solution to obtain disks. In fact all
our models for $Vmax > 120 \, \rm km\,s^{-1}$ --the green line-- (that
is all lines except the orange and magenta ones) coincide in a same
locus for the present time, and reproduce well the observed value
given by \citet{san08} and represented by the red full hexagon.

\begin{figure}
\includegraphics[width=0.35\textwidth,angle=-90]{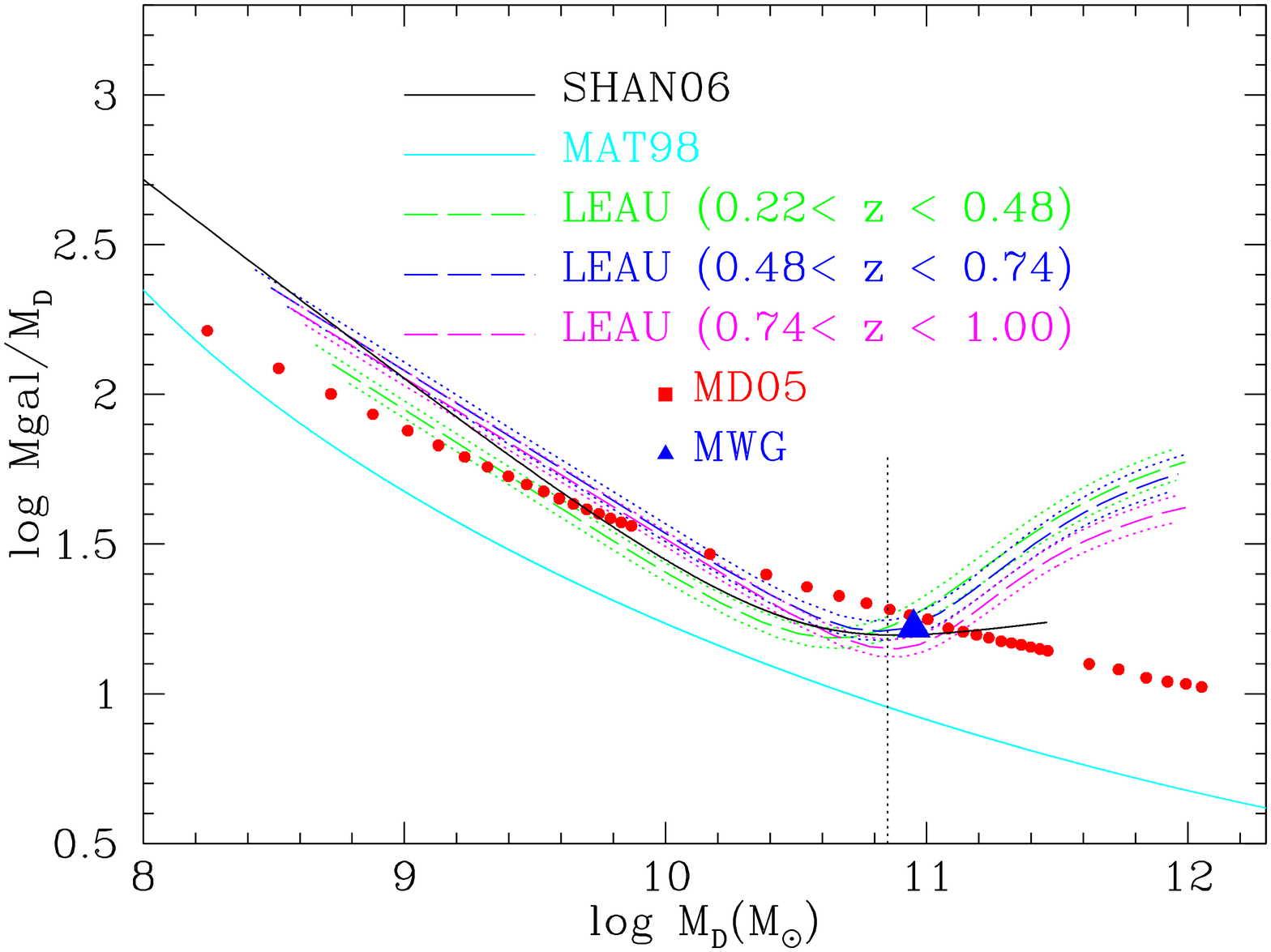}
\caption{The ratio $\frac{Mgal}{M_{D}}$ as a function of the
mass in the disk $M_{D}$. Our models results are the solid red dots. The cyan
and black lines are the results obtained by \citet{mateo98} and
\citet{shan06} from observations of the Local Group of galaxies and
from halo data, respectively.  Magenta, blue and green dashed lines are
results from \citet{leau10} for different ranges of redshift as labeled.
The MWG point is represented by a blue full triangle and the dotted black line
marks the limit between disk and spheroidal galaxies from \citet{gon13}.}
\label{md_mt}
\end{figure}

As a consequence of this infall of gas scenario, the disk is
formed. The proportion of mass in the disk compared with the total
dynamical mass of the galaxy is, as expected, dependent on this total
mass. In Fig.~\ref{md_mt} we show the fraction ${Mgal}/{M_{D}}$, where
$M_{D}$ is the mass in the disk resulting from the applied collapse
time scale prescriptions, as a function of the final mass in the
disk. These results, shown as red points, are compared with the line
obtained by \citet{mateo98} for galaxies in the Local Group, solid
cyan line, and with the ratio by \citet{shan06} calculated through the
halo and the stellar mass distributions in galaxies, solid black
line. We also plot the results obtained by \citet{leau10} from
cosmological simulations for three different ranges in redshift, such
as labeled in the figure. Our results have a similar slope to the one
from \citet{mateo98}, but the absolute value given by this author is
slightly lower, which is probably due to a different value $M_{*}/L$
to transform the observations (luminosities) in stellar masses.  Our
results are close to the ones predicted by \citet{shan06} and
\citet{leau10} obtained with different techniques.  These authors find
in both cases an increase for high disk masses which, obviously, is
not apparent in our models, since we have assumed a continuous
dependence of the collapse time scale with the dynamical mass.
\citet{shan06} analyzed the luminosity function of halos and determine
the relation with the mass formed there, while \citet{leau10} compute
cosmological simulations and obtain the relation between the dynamical
mass and the final mass in their disks. As we say before, disks
obtained in simulations are smaller than observed what increases the
ratio $Mgal/M_{D}$. Moreover the change of slope in these curves
defines the limit in which the elliptical galaxies begin to appear,
shown by the dotted black line as given by \citet{gon13}.  Therefore
it is possible that these authors include some spheroids and galaxies
SO in their calculations which are not computed in our models.  It is
necessary say, however, that the MWG value in this plot is slightly
above this limit, and just where the \citet{shan06} line change the
slope. Taking into account that more massive than MWG there exist,
maybe this limits is not totally correct, and that, instead a sharp
cut, there is a mix of galaxies in this zone of the plot.

\subsection{The relation of the SFR with the molecular gas}

\begin{figure}
\includegraphics[width=0.5\textwidth,angle=0]{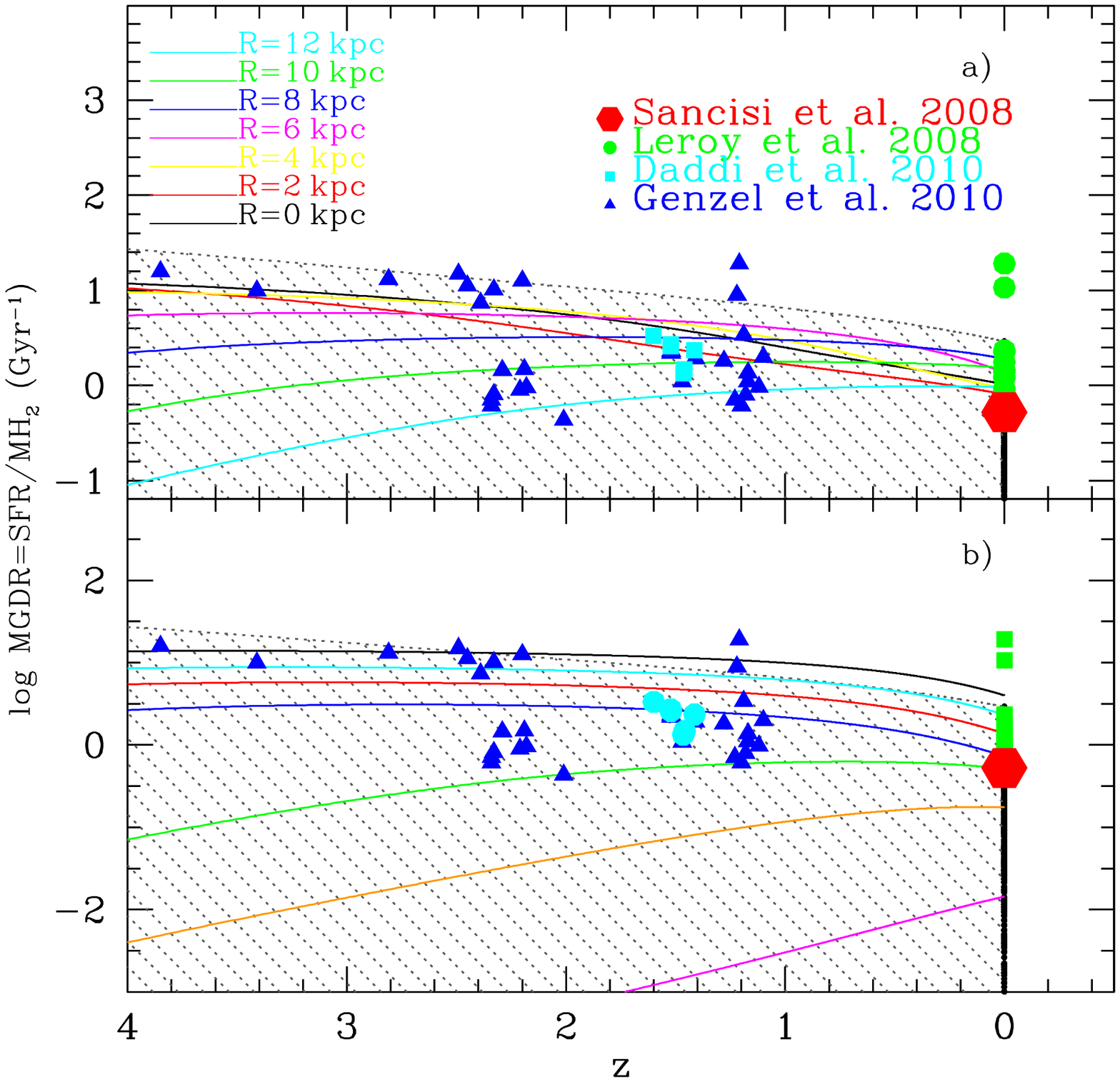}
\caption{The efficiency to form stars from molecular gas,
$SFR/M H_{2}$, in logarithmic scale, as a function of redshift
$z$ for our grid of models. The evolution with redshift for all radial
regions and galaxies are shown by the shaded zone while solid black
points represent the grid results for $z=0$. Upper panel: Solid black,
red, yellow, magenta, blue, green and cyan lines, correspond to radial
regions located at 0, 2, 4, 6, 8, 10 and 12\,kpc of the Galactic
center in the MWG model. Bottom panel: Solid black, cyan, blue, red,
green, orange, and magenta lines correspond to galaxies of different
dynamical masses and efficiencies (see Table~\ref{examples}).  In both
panels observations at intermediate-high redshift are from
\citet{dad10,gen10}, shown as cyan dots and blue triangles,
respectively, while the green squares refers to the local Universe
data obtained by \citet{ler08}. The red star marks the average given
by \citet{san08}.}
\label{mgdr}
\end{figure}

Since the formation of molecular gas is a characteristics which
differentiates our model from other chemical evolution models in the
literature, we would like to check if our resulting star formation is
in agreement with observations.  We compare the efficiency to form
stars from the gas in phase $H_{2}$, measured as $SFR/M H_{2}$, 
with data in Fig.~\ref{mgdr}.  In the upper panel we show the results for a
galaxy like MWG, where each colored line represents a different radial
region: Solid red, yellow, magenta, blue, green and cyan lines,
correspond to radial regions located at 2, 4, 6, 8, 10 and 12\,kpc of
the Galactic center in the MWG model.  In the bottom panel solid black,
cyan, red, blue, green, orange, and magenta lines correspond to the
galaxies of different dynamical masses and efficiencies given in
Table~\ref{examples} and taken as examples. Observations at
intermediate-high redshift by \citet{dad10,gen10}, are shown as cyan
dots and blue triangles, respectively, while the green squares refers
to the local Universe data obtained by \citet{ler08}. The red star
marks the average given by \citet{san08}.

\subsection{Evolution with redshift of the  radial distributions in disks}

\begin{figure*}
\includegraphics[angle=-90,width=\textwidth]{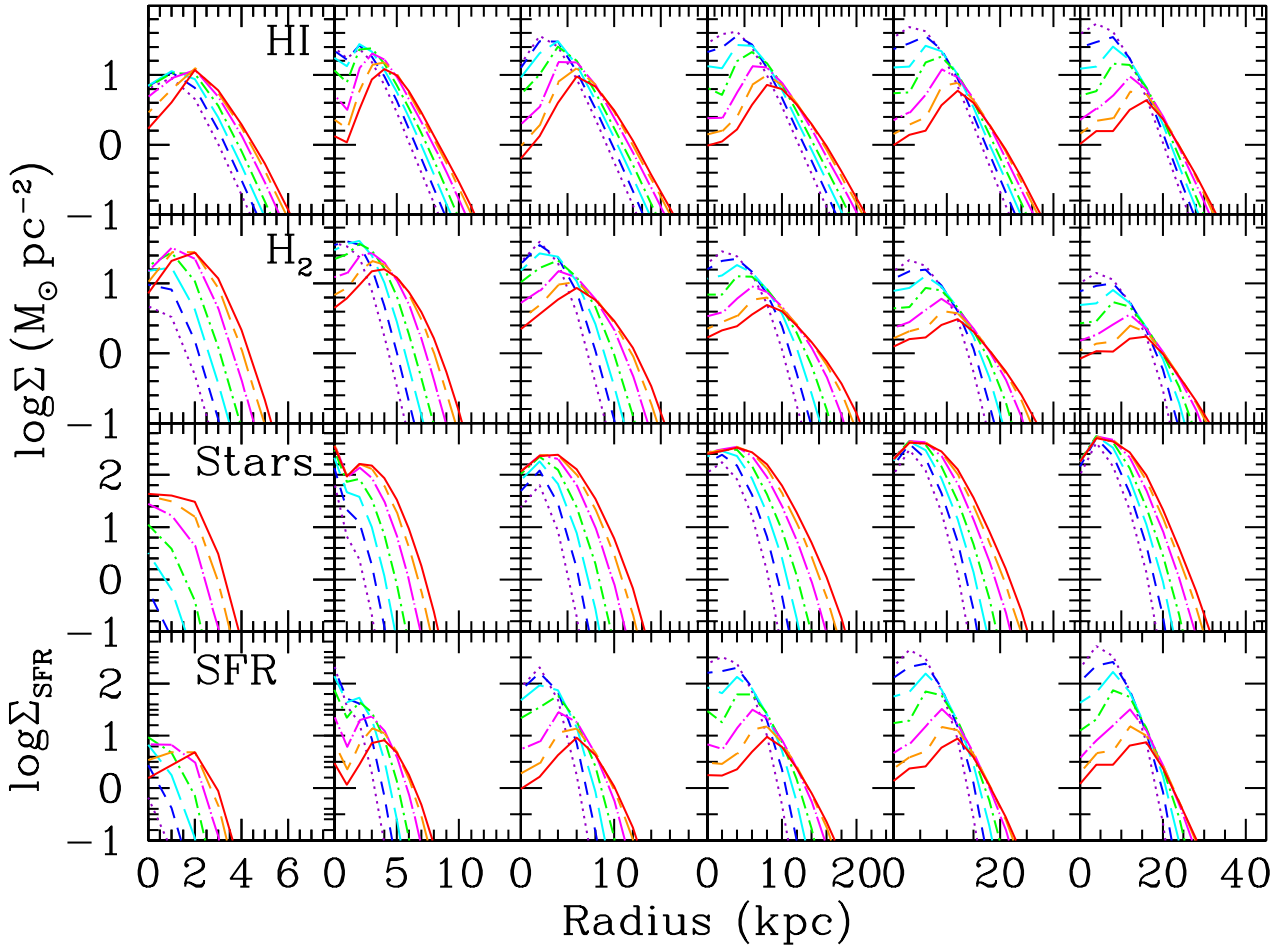}
\caption{Evolution with redshift of the surface density of diffuse
gas, $\Sigma HI$, molecular gas, $\Sigma H_{2}$, and stellar mass,
$\Sigma_{*}$, in $\rm M_{\odot}\,pc^{-2}$, in top, middle-top, and
middle-bottom rows, and the star formation rate, $\Psi$, in $\rm
M_{\odot}\,Gyr^{-1}\,pc^{-2}$, in the bottom one. All in logarithmic
scale. Each column shows the results for a theoretical galaxy of the
Table~\ref{examples}, from the galaxy, with $Vmax=78\,\rm km\,s^{-1}$,
to the most massive one the bottom for $Vmax=290\,\rm km\,s^{-1}$.
Each line corresponds to a different redshift: purple,
blue,cyan,green,magenta, orange and red for $z=$5, 4, 3, 2, 1, 0.4 and
0, respectively}
\label{den_z}
\end{figure*}

In next figures we show the results corresponding to the seven
theoretical galaxies used as examples and whose characteristics are
given in Table~\ref{examples}.  We have selected 7 galaxies which
simulate galaxies along the Hubble Sequence. They have have different
masses and sizes and we have also selected different efficiencies to
form stars in order to compare with real galaxies.  In next figures we
will show our results for 7 values of evolutionary times or redshifts:
$z=5$, 4, 3, 2, 1, 0.4 and 0, with colors purple, blue, cyan, green,
magenta, orange and red, respectively.  The resulting present time
radial distributions in disks for diffuse and molecular gas, stellar
mass, and star formation rate for the galaxies from
Table~\ref{examples} are shown in Fig.\ref{den_z}. In this figure we
show a galaxy in each column and a different quantity in each
row. Thus top, top-middle and bottom-middle panels show the diffuse
gas, the molecular gas and the stellar surface densities, respectively,
all in $\rm M_{\odot}\,pc^{-2}$ units and in logarithmic scale, 
while the bottom panel corresponds to the surface density of the star
formation rate in $\rm M_{\odot}\,Gyr^{-1}\,pc^{-2}$.

The radial distributions of diffuse gas density $\Sigma_{HI}$ show a
maximum in the disk. The radius of the maximum is near the center of
the galaxy for $z=5$ and move outwards with the evolution, reaching a
radius $\sim 2/3\,Ropt$\,kpc for the present, that is 1.3 times $Rc$,
and $\sim 2$ the effective radius in mass or radius enclosed the half
of the stellar mass of the disk (see next section).  The density
$\Sigma_{HI}$ in this maximum reaches values $\sim 50\,\rm
M_{\odot}\,pc^{-2}$ in early times or high redshift. For the present,
the maximum density is $\sim 10\,\rm M_{\odot}\,pc^{-2}$, very similar
in most of theoretical galaxies. These radial distributions reproduce
very well the observations of HI.  The radial distribution for regions
beyond this point shows an exponential decreasing with slopes flatter
now that in the high-redshift distributions.

The radial distributions of molecular gas density $\Sigma_{H_{2}}$
show basically the same behavior, with a maximum in the disks too. In
each galaxy, however, this maximum is located slightly closer to the
center that the one from the HI distribution.  They show an
exponential function too, after this maximum. These radial
distributions seems more an exponential shape with a flattening at the
center that the ones from HI which show a clearer maximum, mainly for
the least massive galaxies.

The stellar profiles, $\Sigma_{*}$, show the classical exponential
disks. The size of these stellar disks, and the scale length of these
exponential functions are in agreement with observations.  However
they present a decreasing or flattening in the inner regions in most
cases, although some of them have a abrupt increase just in the
center.  It is necessary remain that these models are calculated to
model spiral disks, and the bulges are added by hand without any
density radial profile.  Probably this produces a behavior not totally
consistent with observations at the center of galaxies.

In these three panels we show the results only for densities higher
than 0.1\,$\rm M_{\odot}\,pc^{-2}$, which corresponds to a atomic density
$n\sim0.1\,\rm cm^{-2}$ which we consider a lower limit for observations.

The star formation radial distributions show similar shapes than
$\Sigma_{H_{2}}$ and $\Sigma_{*}$ with similar decreasing in the inner
regions of disks. Although these decreases have been observed in a
large number of spiral disks for $\Sigma_{H_{2}}$ and SFR
distributions \citep{mar01,nis01,reg01}, we think, however, that they are
stronger than observed.

From all panels we may say that spiral disks would have a more
compact appearance at high redshift with higher values maximum and
smaller physical sizes, as correspond to a inside-out disk formation
as assumed. The present radial distributions show flatter shapes, with
smaller values in the maximum and in the inner disks and higher
densities in the outer regions.

The star formation is around 2 order of magnitude larger at $z=5$ than
now for the massive galaxies, in agreement with estimated of the star
formation in the Universe \citep{gla99} while for the smaller galaxies
the maximum value is low and very similar then than now.

In Fig.~\ref{lowmass}, we show the same four radial distributions for
the lowest mass galaxy in our grid with a maximum rotation velocity of
$Vmax\sim 50\,\rm km\,s^{-1}$. In this case, and taking into account
the limit of the observational techniques, we see that the galaxy will
be only detected in the region around 1\,kpc and only for redshift
$z<2$ since for earlier times than this the galaxy will be undetected.

\subsection{The half mass radii}

\begin{figure}
\includegraphics[width=0.35\textwidth,angle=-90]{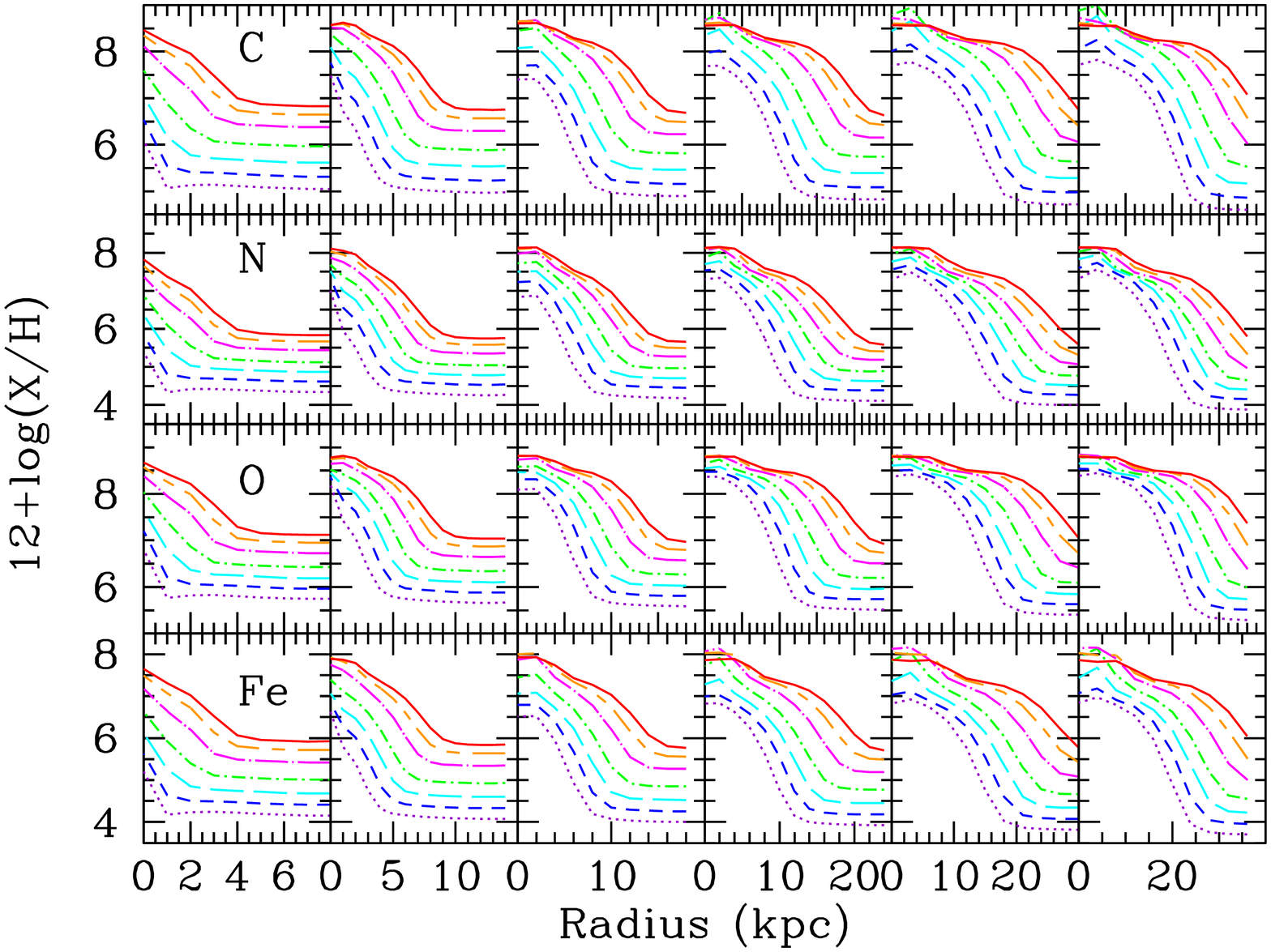}
\caption{The evolution of the half mass radii, $Reff_{mass}$, as a
function of the redshift $z$ for models with efficiencies sets from
$nt=$1 to 9 and for mass distribution numbers from 10 to 44 (see
Tables 1 and 2 from MD) corresponding to maximum rotation velocities
in the range [80-400] $\rm km s^{-1}$.  Black, gray, green, red,
orange,magenta, purple, cyan and blue dots correspond to $nt=$ 1, 2,
3, 4, 5, 6, 7, 8, and 9 respectively.}
\label{reff}
\end{figure}

Since we know the mass of the disk and the corresponding one to each
phase we may follow the increase of the stellar mass in each radial
region and calculate for each evolutionary time o redshift the radius
for which the stellar mass is the half of the total of the mass in
stars in the galaxy, $Reff_{mass}$.  This radius obviously will evolve
with time and since we have assume an in-out scenario of disk
formation, it must increase with it. We show this evolution in
Fig.~\ref{reff}, where we plot with different color the evolution of
different efficiencies models: black, gray, green, red,
orange,magenta, purple, cyan and blue for sets $nt=$1 to 9,
respectively, for mass distributions from number 10 to 44, which
correspond to maximum rotation velocities in the range $Vmax \sim 80$
to 400 $\rm km\,s^{-1}$.  The lowest mass galaxies, with $Vmax <
78\,\rm km\,s^{-1}$ (or those for which efficiencies correspond to
$nt$=10) are not shown since, as we will show in Fig.~\ref{lowmass},
they only would have a visible central region. It is evident that the
effective radius increases, being smaller to high redshift, this
increase begins later for the later types of smaller efficiencies than
for the earlier ones or with high efficiencies.  Blue and cyan points
are below 1\, kpc for redshifts higher than 1, while the others begin
to increase already at $z=$5-6. Moreover a change of slope with a
abrupt increase in the size occurs at $z\sim 3$ until $z\sim 2$, for $
nt< 5$, when the star formation suffers its maximum value in most of
these galaxies which is agreement with the data. Intermediate
galaxies, as magenta and purple dots, which correspond to irregular
galaxies show this change of the slope at $z<2$.

\subsection{Evolution of elemental abundances}

The resulting elemental abundances are given in Table~\ref{abun}. For each 
type of efficiencies $nt$ and mass distribution $dis$, columns 1 and 2, we give the time $t$ in Gyr in column 3,
the corresponding redshift $z$ in column 4, and the Radius in Kpc in column 5.
The elemental abundances for H, D, $^{3}$He, $^{4}$He$, ^{12}$C, $^{13}$C,  N, O, Ne, Mg, Si, S, Ca and Fe  are in columns 6 to 19 as
fraction in mass.

\begin{table*}
\begin{center}
\caption{Evolution of elemental abundances along the time/redshift for the grid of models. We show as example the results
for the present time of a MWG-like model ($nt=4$, $dis=28$). The complete table will be available in electronic format.}
\begin{tabular}{ccccccccc}
\hline
$nt$  & $dis$ &   $t$ & $z$ &  $Radius$ & H & D & $^{3}$He & $^{4}$He \\
        &           &  Gyr &      &    kpc    & & &  \\
\hline
4 & 28 & 1.3201e+01 &  0.00 &   0  &  7.1000e-01  &  7.4918e-08 &   5.2929e-04  &  2.7149e-01 \\
4 & 28 & 1.3201e+01 &  0.00 &   2  &  7.1020e-01  &  3.7168e-08 &   5.2559e-04  &  2.7085e-01 \\
4 & 28 & 1.3201e+01 &  0.00 &   4  &  7.0751e-01  &  1.4105e-06 &   4.9819e-04  &  2.7373e-01 \\
4 & 28 & 1.3201e+01 &  0.00 &   6  &  7.2710e-01  &  2.9584e-05 &   2.3378e-04  &  2.5935e-01 \\
4 & 28 & 1.3201e+01 &  0.00 &   8  &  7.4243e-01  &  4.6899e-05 &   1.0063e-04  &  2.4818e-01 \\
4 & 28 & 1.3201e+01 &  0.00 &  10  &  7.4664e-01  &  5.1180e-05 &   7.1091e-05  &  2.4520e-01 \\
4 & 28 & 1.3201e+01 &  0.00 &  12  &  7.4958e-01  &  5.3626e-05 &   5.6599e-05  &  2.4321e-01 \\
4 & 28 & 1.3201e+01 &  0.00 &  14  &  7.5423e-01  &  5.7213e-05 &   4.2096e-05  &  2.4017e-01 \\
4 & 28 & 1.3201e+01 &  0.00 &  16  &  7.6037e-01  &  6.2123e-05 &   2.7653e-05  &  2.3619e-01 \\
4 & 28 & 1.3201e+01 &  0.00 &  18  &  7.6558e-01  &  6.6386e-05 &   1.7397e-05  &  2.3283e-01 \\
4 & 28 & 1.3201e+01 &  0.00 &  20  &  7.6841e-01  &  6.8681e-05 &   1.2615e-05  &  2.3100e-01 \\
4 & 28 & 1.3201e+01 &  0.00 &  22  &  7.6931e-01  &  6.9386e-05 &   1.1281e-05  &  2.3043e-01 \\
4 & 28 & 1.3201e+01 &  0.00 &  24  &  7.6947e-01  &  6.9516e-05 &   1.1048e-05  &  2.3032e-01 \\
\hline
\label{abun}
\end{tabular}
\end{center}
\end{table*}

\begin{table*}
\addtocounter{table}{-1}
\begin{center}
\caption{Cont. Evolution of elemental abundances along the time/redshift for the grid of models}
\begin{tabular}{cccccccccc}
\hline
 $^{12}$C & $^{13}$C & N & O & Ne & Mg & Si & S & Ca & Fe \\
     &                &    &     &      &      &      &    &     &    \\ 
\hline
 3.1109e-03 &   5.6767e-05 &   1.3390e-03 &   6.9834e-03 &   1.1736e-03 &   3.2207e-04 &   1.2311e-03 &   6.2757e-04 &   8.8136e-05 &   2.8875e-03\\
 3.1888e-03 &   5.8285e-05 &   1.3916e-03 &   7.1115e-03 &   1.1987e-03 &   3.3036e-04 &   1.2594e-03 &   6.4239e-04 &   9.0264e-05 &   2.9750e-03\\
 3.1561e-03 &   5.3970e-05 &   1.2723e-03 &   7.0513e-03 &   1.1641e-03 &   3.1662e-04 &   1.2872e-03 &   6.5643e-04 &   9.1892e-05 &   3.0407e-03\\
 2.4417e-03 &   2.8054e-05 &   7.2097e-04 &   5.4124e-03 &   8.6307e-04 &   2.2721e-04 &   9.2775e-04 &   4.7042e-04 &   6.5617e-05 &   2.0596e-03\\
 1.7736e-03 &   1.3776e-05 &   4.0653e-04 &   4.0741e-03 &   6.4323e-04 &   1.6486e-04 &   6.0974e-04 &   3.0600e-04 &   4.2665e-05 &   1.2016e-03\\
 1.5606e-03 &   1.0075e-05 &   3.1545e-04 &   3.6797e-03 &   5.7863e-04 &   1.4650e-04 &   5.1733e-04 &   2.5795e-04 &   3.5927e-05 &   9.5008e-04\\
 1.3936e-03 &   7.4918e-06 &   2.3852e-04 &   3.3407e-03 &   5.2112e-04 &   1.3070e-04 &   4.5084e-04 &   2.2362e-04 &   3.1087e-05 &   7.8546e-04\\
 1.0958e-03 &   4.5570e-06 &   1.4171e-04 &   2.6790e-03 &   4.1288e-04 &   1.0266e-04 &   3.4491e-04 &   1.7020e-04 &   2.3619e-05 &   5.6743e-04\\
 6.7707e-04 &   2.1410e-06 &   6.2137e-05 &   1.6978e-03 &   2.5908e-04 &   6.4119e-05 &   2.0782e-04 &   1.0209e-04 &   1.4164e-05 &   3.2295e-04\\
 3.1234e-04 &   8.0794e-07 &   2.2182e-05 &   8.0723e-04 &   1.2299e-04 &   3.0589e-05 &   9.4069e-05 &   4.6086e-05 &   6.4103e-06 &   1.4037e-04\\
 1.1295e-04 &   2.7003e-07 &   8.2285e-06 &   2.9966e-04 &   4.6226e-05 &   1.1834e-05 &   3.3587e-05 &   1.6495e-05 &   2.3115e-06 &   5.1772e-05\\
 5.0752e-05 &   1.2185e-07 &   4.7315e-06 &   1.3420e-04 &   2.1235e-05 &   5.7565e-06 &   1.4834e-05 &   7.3585e-06 &   1.0446e-06 &   2.6329e-05\\
 3.9167e-05 &   9.4973e-08 &   4.1239e-06 &   1.0244e-04 &   1.6438e-05 &   4.5963e-06 &   1.1325e-05 &   5.6530e-06 &   8.0808e-07 &   2.1822e-05\\
\hline
\end{tabular}
\end{center}
\end{table*}

\begin{figure*}
\includegraphics[angle=-90,width=\textwidth]{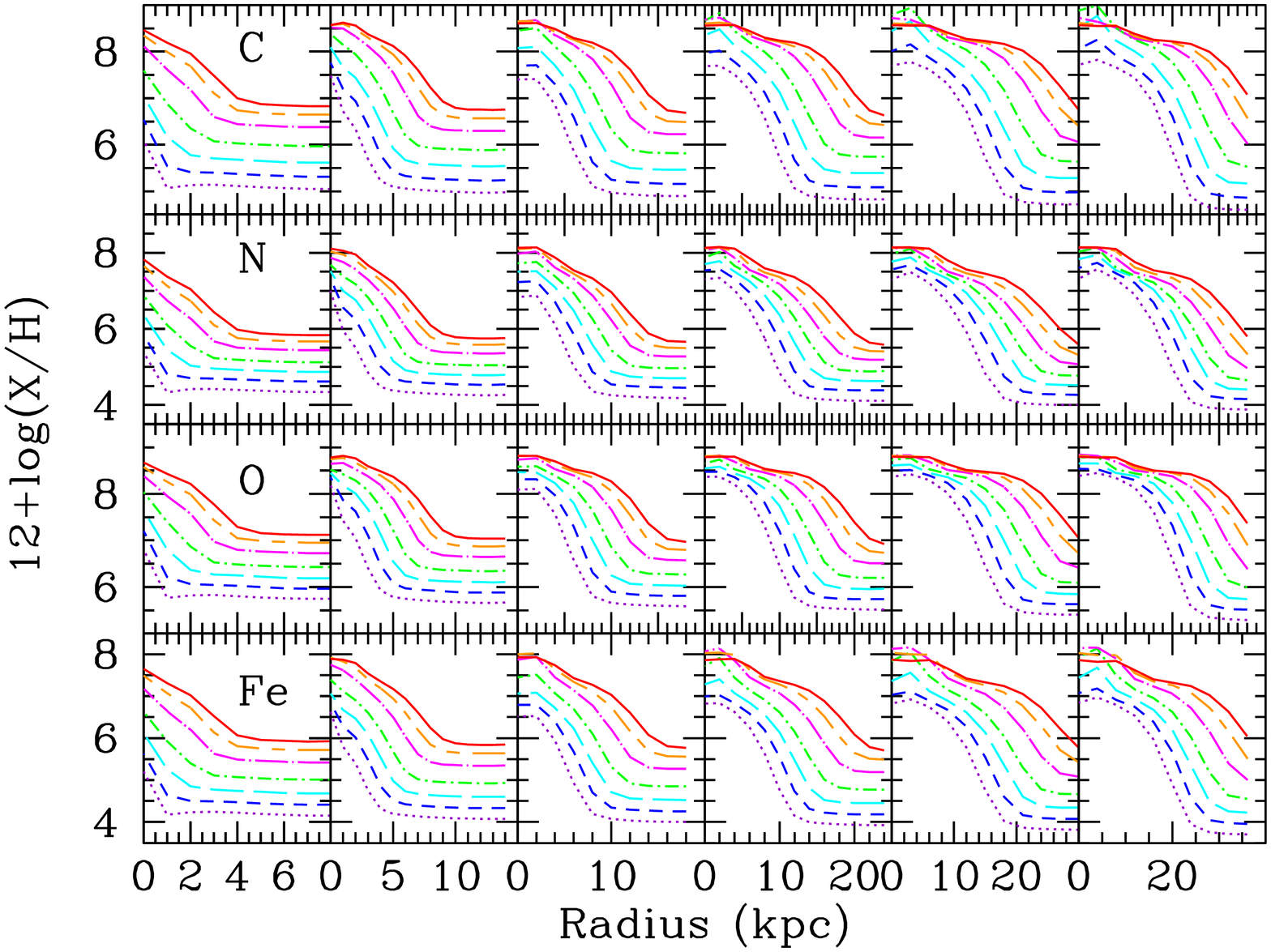}
\caption{Evolution with redshift of radial distributions of elemental
  abundances, C, N, O and Fe from top to bottom Each column shows the
  results for a theoretical galaxy of the Table~\ref{examples}, from
  the galaxy, with $Vmax=78\,\rm km\,s^{-1}$, at left to the most massive
  one with $Vmax=290\,\rm km\,s^{-1}$ at the right.  In each panel
  distributions for 7 redshifts are shown, with the same color coding
  than in Fig.~6}
\label{abun_z}
\end{figure*}

The radial distributions of elemental abundances are shown in
Fig.~\ref{abun_z}. There we have for the same 6 galaxies of
Fig.~\ref{den_z} from the left to the right column, the abundance
evolution for C, N, O and Fe, as $12+\log{(X/H)}$ from top to bottom
panels.  In each panel, as before, we represent the radial
distributions for 7 different redshifts, $z=$ 5, 4, 3, 2, 1, 0.4 and
0, with the same color coding. The well known decreasing from the
inner to the outer regions, called the radial gradients of abundances,
appear in the the final radial distributions for the present time in
agreement with data.  However the slope is not unique in most of
cases. The radial distribution for any element, $12+\log{(X/H)}$,
is not a straight line but a curve which is flatter in the central
region and also in the outer disk in agreement with the most recent
observations \citep{san13}.

The distributions of abundances change with the level of evolution of
a galaxy. A very evolved galaxy, that is, as the one for the most
massive ones, and/or those with the highest efficiencies to form
stars, show flatter radial gradients that those which evolve slowly,
which have steeper distributions.  There exists a saturation level of
abundances, defined by the true yield, and given by the combination of
stars of a certain range of mass and the production of elements of
these stars. For our combination of stellar yields from
\citet{woo95,gav05,gav06} and IMF from \citet{fer90}, this level is
$12+\log{(X/H)}\sim 8.8, 8.0, 9.0$ and 8.2 dex for C, N, O and Fe
respectively. As the galaxy evolves the saturation level is reached in
the outer regions of galaxy, thus flattening the gradient.

\begin{figure*}
\subfigure{\includegraphics[width=0.35\textwidth,angle=-90]{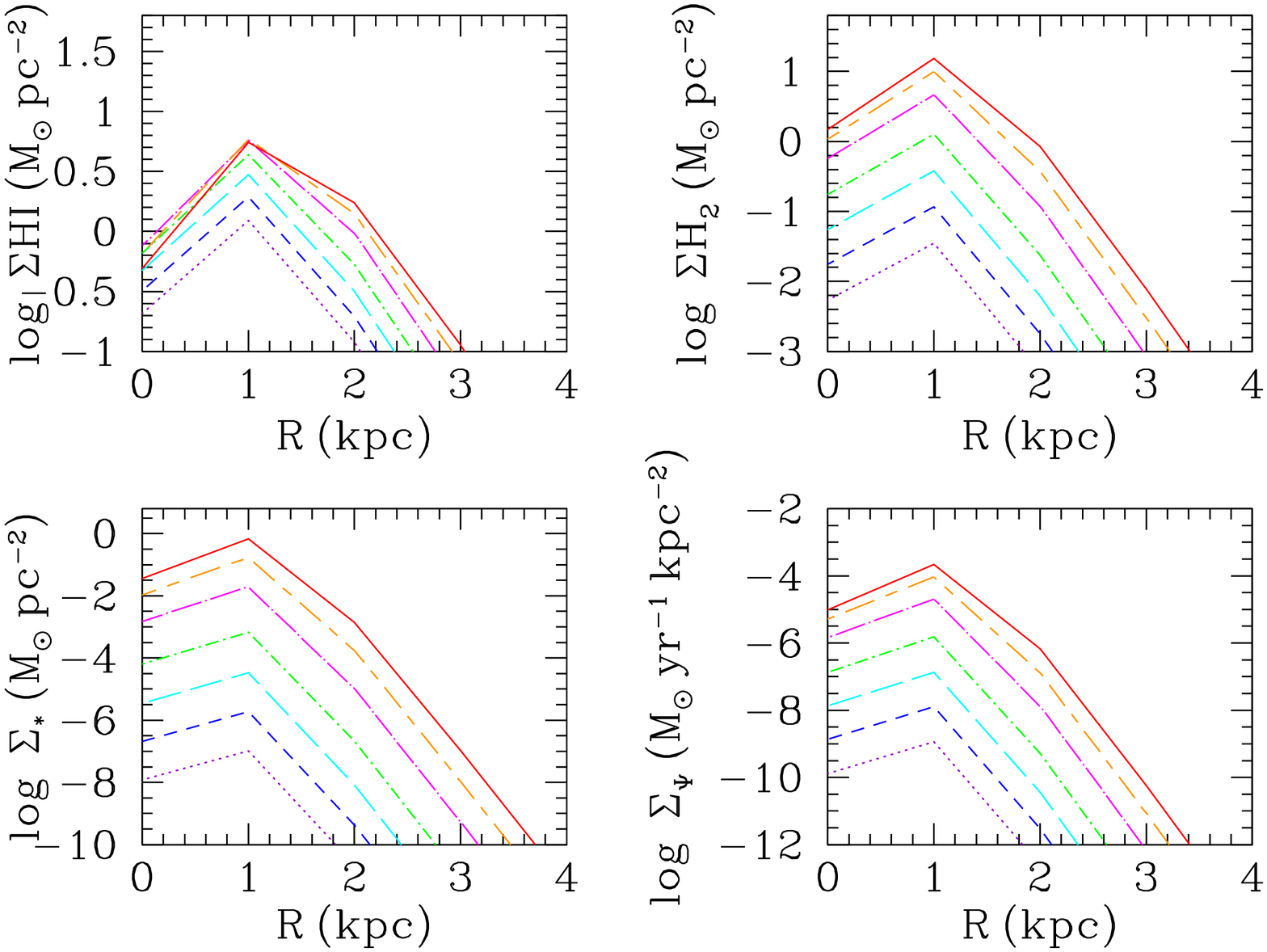}}
\subfigure{\includegraphics[width=0.35\textwidth,angle=-90]{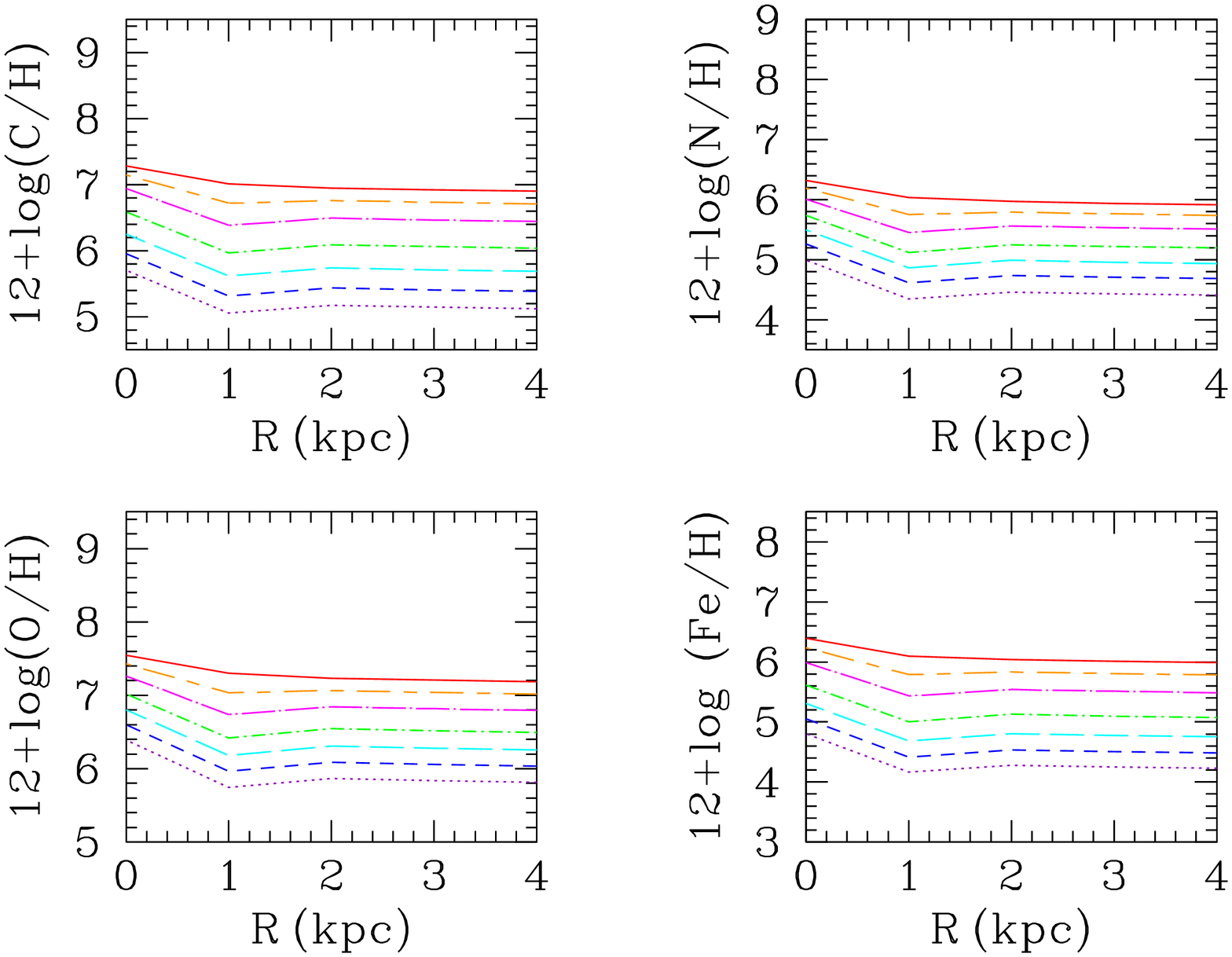}}
\caption{Left. Radial distributions of the surface density of diffuse
gas, $\Sigma HI$, molecular gas, $\Sigma H_{2}$, and stellar mass,
$\Sigma_{*}$, in $\rm M_{\odot}\,pc^{-2}$, and the star formation rate
surface density, $\Sigma_{SFR}$, in $\rm M_{\odot}\,yr^{-1}\,pc^{-2}$, all in
logarithmic scale, for the least massive galaxy of our example table,
with $Vmax=48\,\rm km\,s^{-1}$.  Right. Evolution with redshift of
radial distributions of elemental abundances, C, N, O and Fe.}
\label{lowmass}
\end{figure*}

The radial gradient flattens with the evolutionary time or redshift,
mainly in the most massive galaxies. However it maintains a similar
value for the smallest galaxies. The extension in which the radial
gradient appears, however, changes with time. At the earliest time the
radial gradient appears only for the central regions, until 1\,kpc in
the left column galaxy, while it does until 22\,kpc in the right one,
with a change of slope at around 8\,kpc. For the radial regions out of
this limit, it shows a flat distribution. At the present, this
gradient appears until a more extended radius, 28\,kpc in our most
massive galaxy, while it is only in the inner 4\,kpc in the smallest
one.  If we analyze the results for the lowest mass galaxy in
Fig.~\ref{lowmass}, we see that abundances show a very uniform
distribution along the galactocentric radius for all elements, with a
slight increase at the center for all redshifts that may be considered
as nonexistent within the usual errors bars.

The flat radial gradient in the lowest mass galaxies, as shown in
Fig.~\ref{lowmass}, as this one of the most distant regions of the disk in the
massive galaxies at the highest redshifts, must be considered as a
product of the infall of a gas more rich that this one of the disk. It
is necessary to take into account that the halo is forming stars
too. When the collapse time is longer, as occurs in the outer regions
of disks, there is more time and more gas in the halo to form stars
and increase its abundances. Thus, the gas infalling is, even at a
very low level ($12 +\log{(X/H)}\sim 4-5$), more enriched than the gas of
the disk.

\section{The photometric model description}

It is well known that galaxies have SEDs depending on their
morphological type \citep{col80}.  These SEDs, and other data related
to the stellar phase, are usually analyzed through (evolutionary)
synthesis models \citep[see][ for a recent and updated review about
these models]{conroy13}, based on SSPs created by an instantaneous
burst of star formation (SF).  The synthesis models began calculating
the luminosity (in a broad band filter or as a SED,
$F_{\lambda}(\lambda)$, or by using the spectral absorption indices)
for a generation of stars created simultaneously, (therefore with a
same age, $\tau$ and with a same metallicity $Z$), that is the
so-called Single Stellar Population (SSP).  The evolutionary codes
compute the corresponding colors, surface brightness and/or spectral
absorption indices emitted by a SSP from the sum of spectra of all
stars created and distributed along a Hertzsprung Russell diagram,
weighted with an IMF.  This SED, given $\tau$ and $Z$, is
characteristic of each SSP. This way it is possible to extract some
information of the evolution of galaxies is by using evolutionary
synthesis models in comparison with spectro-photometric
observations. This method has been very useful for the study of
elliptical galaxies, for which was developed, with the hypothesis that
they are practically SSPs, allowed to advance very much in the
knowledge of these objects, determining their age and metallicity with
good accurate \citep{cb91,bc93,bre94,wor94, fioc97,lei99,vaz99}.

Star formation history, however, does not always take place in a
single burst, as occurs in spiral and irregular galaxies where star
formation is continuous or in successive bursts. Since in these
galaxies the star formation does not occurs in a single burst the SSP
SEDs are not good representative of their luminosity.  In this case it
is necessary to perform a convolution of these SEDs with the star
formation history (SFH) of the galaxy, $\Psi(t)$.  Thus, spectral
evolution models of galaxies predict colors and luminosities of a
galaxy as a function of time, as for example \citet[][ hereafter {\sc
galaxev}]{bc03,bc11}, or the ones from \citet{fioc97}, and \citet[][
hereafter \sc{pegase} \rm 1.0 and 2.0, respectively]{bor04}, from the SEDs
calculated for the SSPs and also for some possible combinations of
them by following a given SFH. Usually some hypotheses about the shape
and the intensity of the SFH, are assumed, e.g. an exponentially
decreasing function of time is normally used. However an important
point, usually forgotten when this technique is applied, is that
$S_{\lambda}(\tau,Z(R))= S_{\lambda}(\tau,Z(R,t'))$, that is, the
metallicity changes with time since stars form and die
continuously. It is not clear which $Z$ must be selected at each time
step without knowing this function $Z(t)$. Usually, only one $Z$ is
used for the whole integration which may be an over-simplification.

Besides this fact that most of these models do not compute the
chemical evolution that occurs along the time, (or do it in a very
simple way), the star formation histories are assumed as inputs.  In
our approach we make take advantage from the results of the chemical
evolution models section, which give to us as outputs the SFH and the
AMR, and use them as inputs to compute the SED of each galaxy or
radial region.  For each stellar generation created in the time step
$t'$, a SSP-SED, $S(\tau,Z(R,t))$, from this set is chosen taking into
account its age, $\tau=t-t'$, from the time $t'$ in which it was
created until the present $t$, and the metallicity $Z(R,t')$ reached
by the gas. After convolution with the SFH, $\Psi(R,t)$, the final
SED, $F_{\lambda}(\lambda,R,t)$, is obtained for each region.  This
way in a region of each galaxy, the final SED, $F_{\lambda}$,
corresponds to the light emitted by the successive generations of
stars. It may be calculated as the sum of several SSP SEDs,
$S_{\lambda}$, being weighted by the created stellar mass in each time
step, $\Psi(R,t)$. Thus, for each radial region:

\begin{equation}
F_{\lambda}(R,t)=\int_{0}^{t} S_{\lambda}(\tau,Z(R,t'))\Psi(R,t')dt',
\label{Flujo}
\end{equation}
where $\tau=t-t'$.

The set of SSP's SEDs, $S_{\lambda}(\tau,Z)$ used are those from the
{\sc popstar} evolutionary synthesis code \citep{mol09}. The
isochrones used in that work are those from \cite{bgs98} for
six different metallicities: Z $=$ 0.0001, 0.0004, 0.004, 0.008, 0.02
and 0.05, updated and computed for that particular piece of work.
The age coverage is from $\log{\tau}=$ 5.00 to 10.30 with a variable
time resolution which is $\Delta(\log{\tau})=0.01$ in the youngest
stellar ages, doing a total of 106 ages.  The WC and WN stars are
identified in each isochrone according to their
surface abundances. The grid is computed for six different
IMF's. Here we have used the set calculated with the IMF from
\citet{fer90} to be consistent with the one used in the chemical
evolution models grid.  To each star in the HR diagram a stellar model
is assigned based in the effective temperature and in gravity. Stellar
atmosphere models are taken from \citet{lcb97}, due to its expansive
coverage in effective temperature, gravity, and metallicities, for
stars with Teff $\leq 25000$K.  For O, B, and WR stars, the NLTE
blanketed models of \citet{snc02} (for metallicities Z $=0.001$,
0.004, 0.008, 0.02, and 0.04) are used. There are 110 models for O-B
stars, calculated by \citet{phl01}, with 25000\,K $<$ Teff $\leq
51500$\,K and $2.95 \leq \log{g} \leq 4.00$, and 120 models for WR
stars (60 WN and 60 WC), from \citet{hm98}, with 30000\,K $\leq
T^{*} \leq 120000$\,K and $1.3\,R_{\odot}\leq R^{*}\leq
20.3\,R_{\odot}$ for WN, and with $ 40000\,K \leq T^{*} \leq
140000$\,K and $ 0.8\,R_{\odot}\leq R^{*}\leq 9.3 \,R_{\odot}$ for
WC. T$^{*}$ and R$^{*}$ are the temperature and the radius at a
Roseland optical depth of 10. The assignment of the appropriate WR
model is consistently made by using the relationships between
opacity, mass loss, and velocity wind, as described in \cite{mol09}. For
post-AGB and planetary nebulae (PN) with T$_{\rm eff}$ between
50000\,K and 220000\,K, the NLTE models from \citet{rau03} are
taken. For higher temperatures, {\sc PopStar} uses black-bodies. The
use of these latter models modifies the resulting intermediate age
SEDs.  The range of wavelengths s the same that the one from
\cite{lcb97}, from 91\,\AA\ to 160,$\mu$~m.

\section{Spectro-photometric Results}
\subsection{Spectral Energy distributions}

\begin{figure*}
\centering
\subfigure{\includegraphics[width=0.45\textwidth,angle=0]{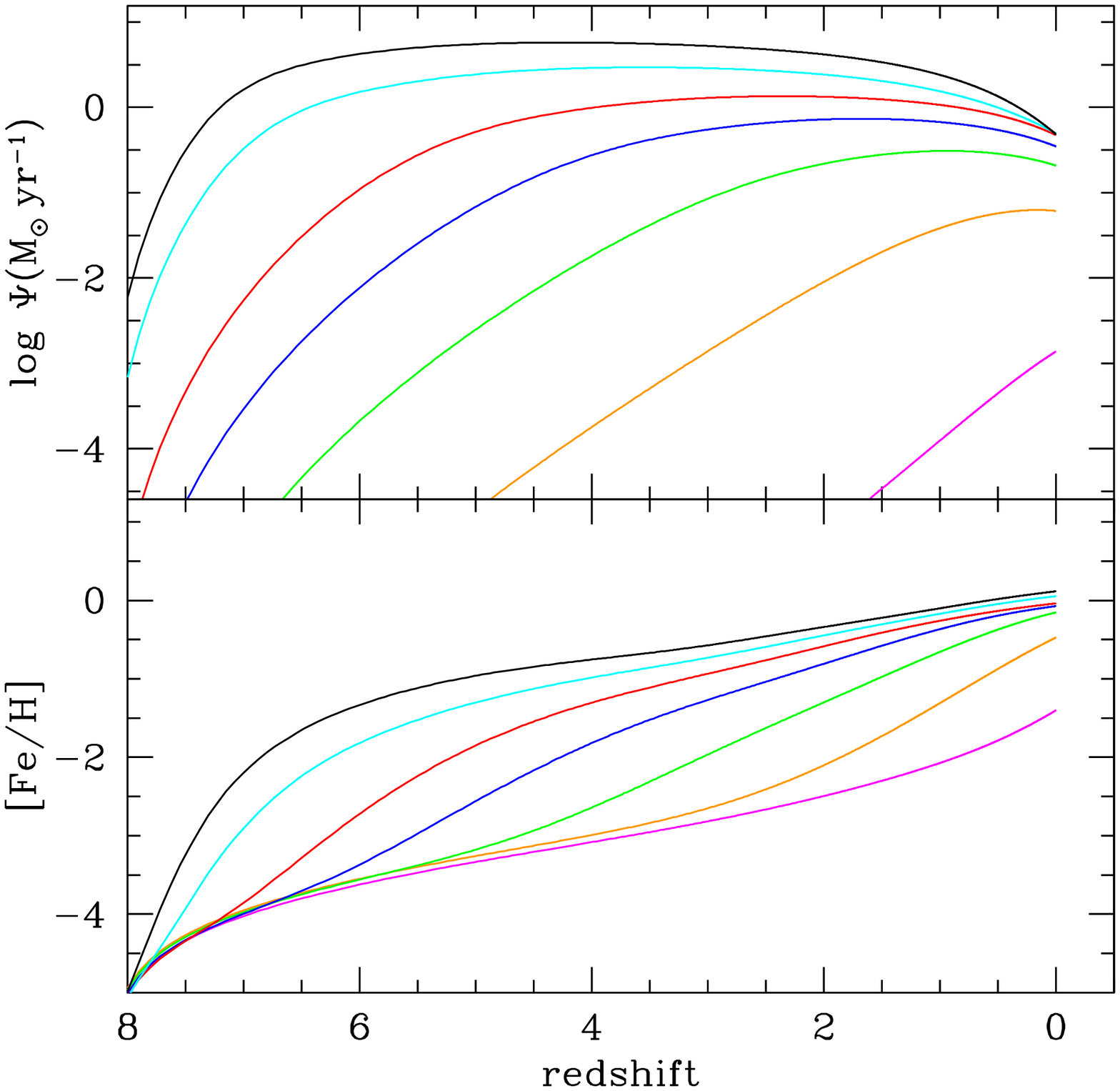}}
\subfigure{\includegraphics[width=0.45\textwidth,angle=0]{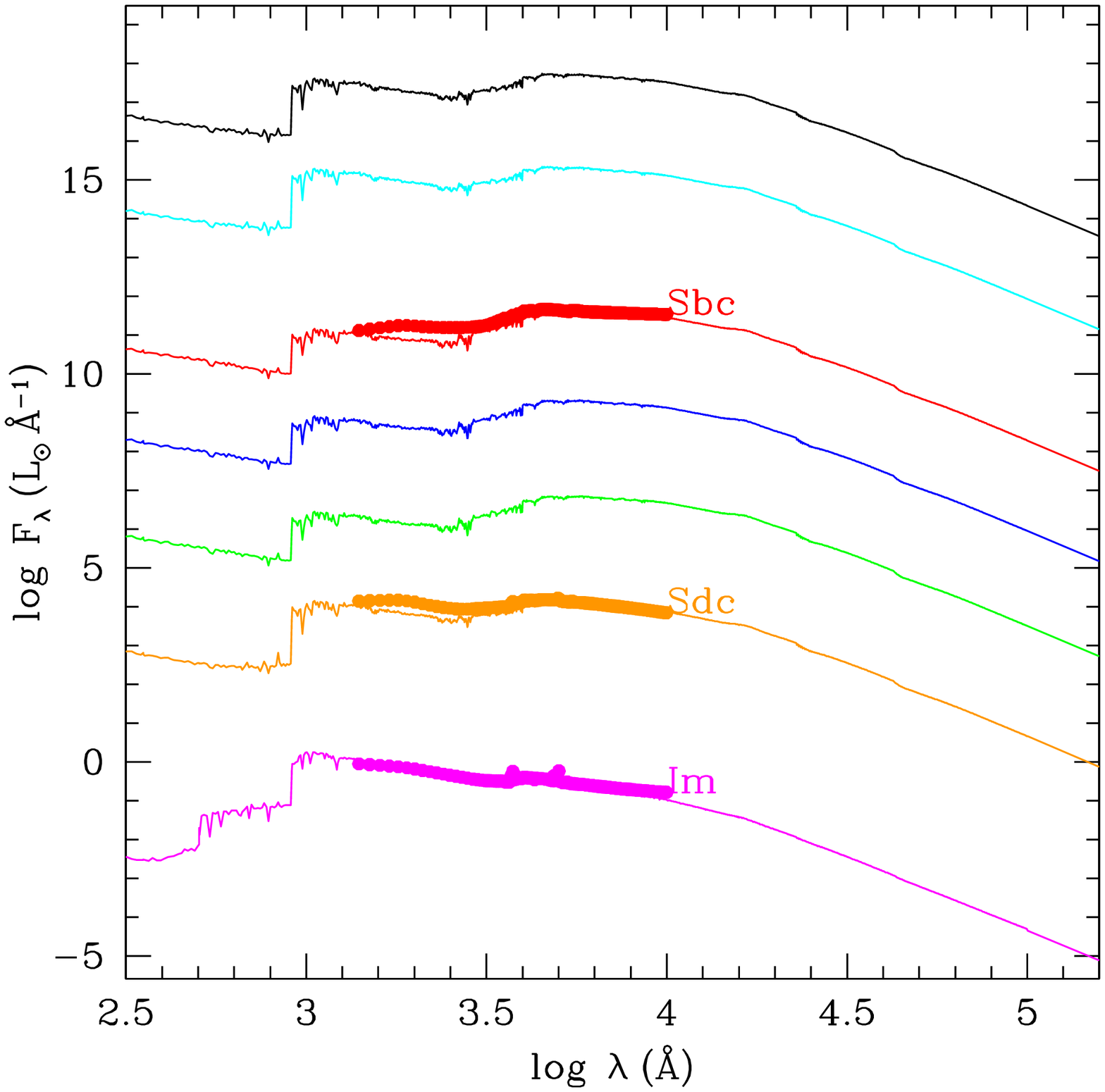}} 
\caption{Left. The star formation $\Psi(t)$, and iron enrichment
$[Fe/H](t)$ evolution with redshift $z$ for the radial regions located
at $Rc$ of models of Table~\ref{examples}. Right. The resulting
spectral energy distributions, $F_{\lambda}(\lambda,t=13.2\,\rm Gyr)$
obtained for same radial region and galaxy model compared with the
fiducial templates from \citet{col80}.}
\label{sfr_feh_sed}
\end{figure*}

As an example of the technique described in the above section, we show
the star formation history $\Psi(t)$ and the age metallicity
distributions $Z(t)$, as $[Fe/H](t)$, for the characteristic radius,
$Rc$, regions of the galaxies of Table~\ref{examples} in
Fig.~\ref{sfr_feh_sed}.  By using these histories we obtain the
resulting $F_{\lambda}(\lambda,t)$ for these regions which reproduce
reasonably well the SEDs of the sampled galaxies.  We have compared
our resulting spectra after a time 13.2 Gyr of evolution for the
models for galaxies from Table~\ref{examples} with the known templates
for the different morphological types from \cite{col80,buz05,bos03},
such as can be seen in Fig.~\ref{sfr_feh_sed} where we compared our
results with \citet{buz05}. We have represented galaxies ordered by
the galactic mass, with the most massive in the top of the graph and
each SED is shifted 2 dex compared with the previous one in sake of
clarity of the figure. There are some differences that probably are
related with the fact that we are comparing a whole galaxy SED with
the one modeled for a given radial region. With this we try to check
that our resulting SEDs are in reasonable agreement with observations.
A more detailed comparison with data for different radial regions of
disks using both types of information, this one proceeding from the
gas, the present state of disks, and the one coming from the stellar
populations and represented by the brightness profiles in different
broad bands, is beyond the scope of this work and will be the object of
a next publication.

\begin{figure*}
\centering
\subfigure{\includegraphics[width=0.45\textwidth,angle=0]{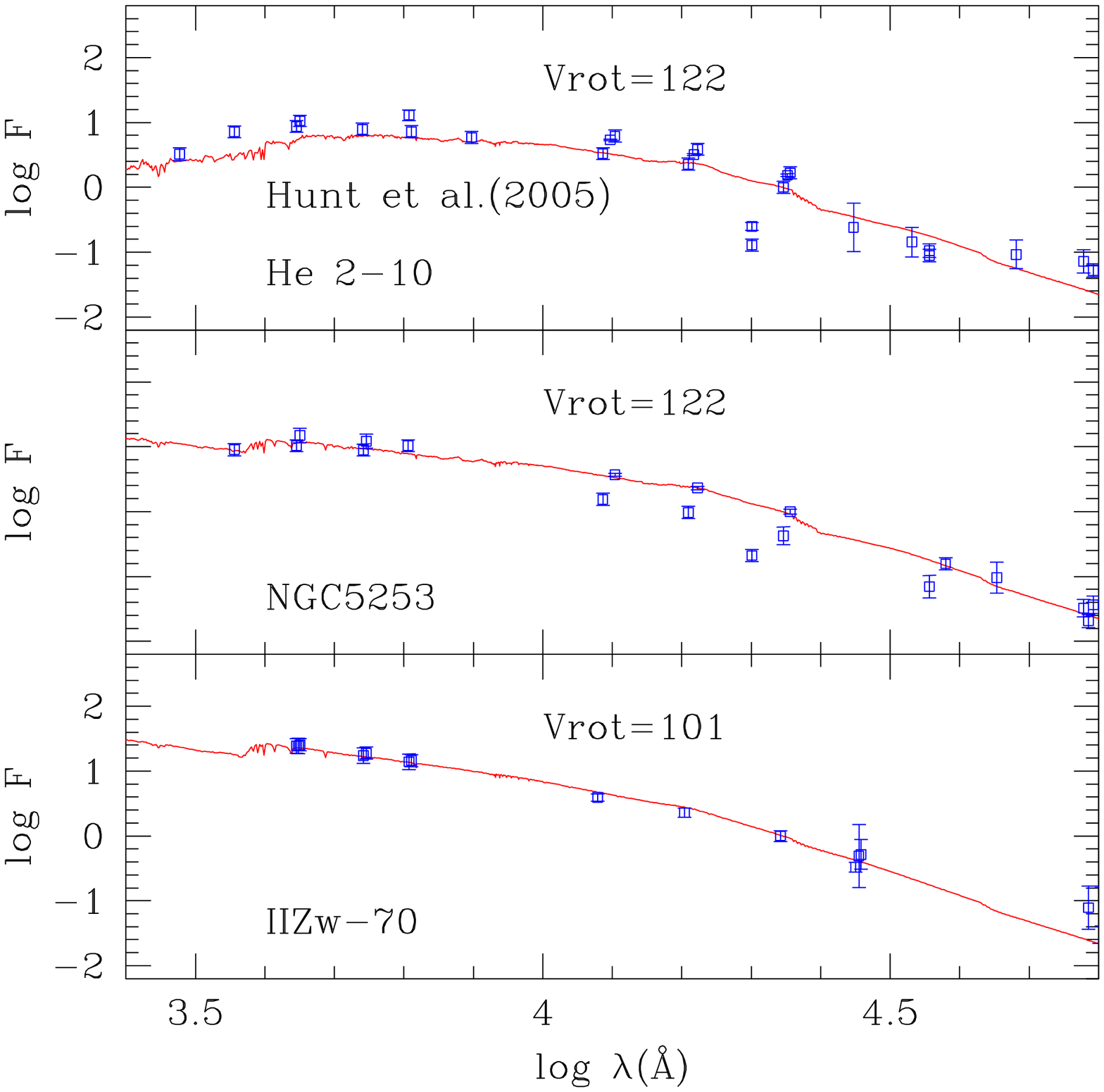}}
\subfigure{\includegraphics[width=0.45\textwidth,angle=0]{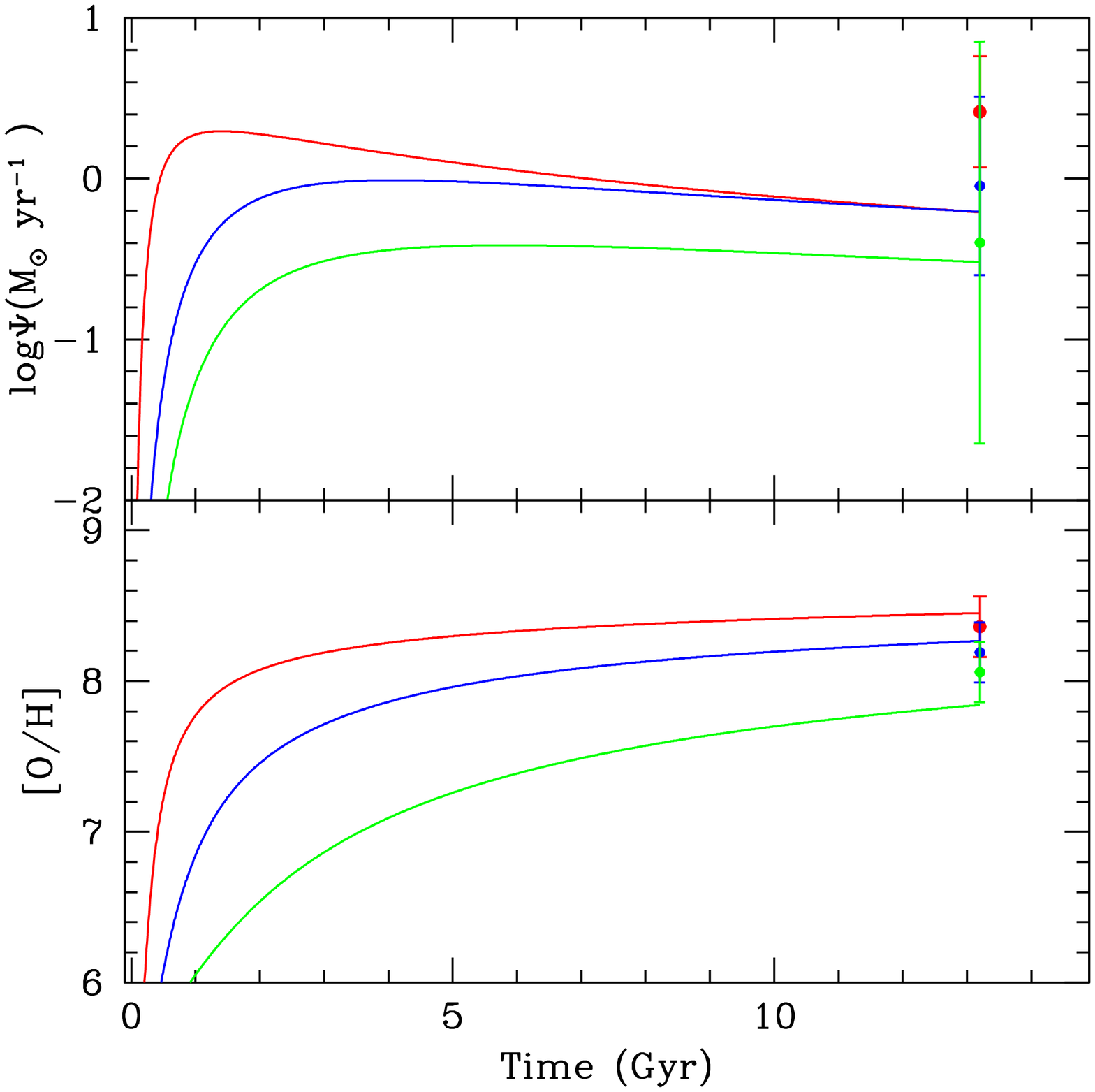}}
\caption{Left. The resulting spectral energy distributions --red solid
lines- obtained to reproduce the observations from \citet{hunt05}
--blue open squares- for three BCD galaxies.  Right. The star
formation history and age-metallicity relation with which we obtain
the SEDs which best fit to the observed spectra.}
\label{sp_bcd}
\end{figure*}

In order to use our model grid, we may therefore to select the best
model able to fit a given observed SED and then see if the
corresponding SFH and the AMR of this model are also able to reproduce
the present time observational data of SFR and metallicity of the
galaxy or of the radial region.  We have a SED for each radial region,
so we may compare the information coming from different radial regions
with our models.  When only a spectrum is observed for one galaxy, it
is possible to compare with the characteristic radius region or to add
all spectra of a galaxy. For dwarf galaxies both methods give
equivalent results since only central regions have suffered star
formation enough to create visible stellar populations.  We show a
example of this method in Fig.~\ref{sp_bcd}, left panel, where three
SEDs from \cite{hunt05} of BCD galaxies are compared with the
characteristic region spectrum of the best model chosen for each one
of them. In the right panel of the same figure we show the
corresponding SFH, $\Psi(t)$, and AMR, $[Fe/H](t)$, with which the
modeled SEDs were computed. The final values are within the error bars
of observations for these galaxies, compiled by the same authors.
Since each SED is well fitted and, simultaneously, the corresponding
present--time data of the galaxy by the same chemical evolution model,
we may be confident that these SFH and AMR give to us a reliable
characterization of the evolutionary history of each galaxy.

It is clear that spiral and irregular galaxies are systems more
complex than those represented by SSP's, and that, in particular,
their chemical evolution must be taken into account for a precise
interpretation of the spectro-photometric data. On the plus side for
these objects the gas phase data are also available and may be used as
constraint.  What is required then is to determine the possible
evolutionary paths followed by a galaxy that arrive at the observed
present state, while, simultaneously, reproducing the average
photometric properties defined by the possible underlying stellar
populations.

\subsection{Broad-band magnitudes and Color-color diagrams}

Once the SEDs obtained for all times/redshifts and radial regions of
our whole grid of models, we may calculated the magnitudes in the
usual broad-band filters. In this case we have computed these ones in
the Johnson and SDSS/SLOAN systems by following the prescriptions
given in \citet{gir02,gir04}.  For Johnson-Cousins-Glass magnitudes
$UV1$, $UV2$, $U$, $B$, $V$, $R$, $I$, $J$, $H$, $K$, and $L$ are
computed using the definition suitable for photon counting devices
\citep{gir02}. Absolute magnitudes in the AB-SDSS photometric system have been
calculated following \citet*{gir04} and \citet{smith02}.
See more details about this in the refereed works or in
\citet{mol09} where the magnitudes were obtained for the SSP-SEDs.

The results are absolute magnitudes in the rest-frame of an observed
located a distance $d=10\,\rm pc$. Therefore we have not used the distance at
which a galaxy a given redshift must be, neither redshift of the
wavelength. These effects must be take into account when observational
apparent magnitudes are calculated. In that case it is necessary to
calculate the decreasing of the flux due to the distance, to include
the K-correction and the wavelength redshift.

These absolute magnitudes are given in Table~\ref{mag}, where for each
set of efficiencies, defined by $nt$ in column 1, and for each radial
mass distribution, defined by the number given in Table~1 from MD05,
$dis$, in column 2, we have the evolutionary time, $t$ in Gyr, in
column 3, and the associated redshift $z$, in column 4, the radius of
each radial region, $Radius$ in kpc, in column 5, and the
corresponding disk area, $Area$, in $\rm kpc^{2}$, in column 6. Then,
we have two ultraviolet from the Hubble telescope, and the classical
Johnson system magnitudes, $UV_{1}$, $UV_{2}$, $U$, $B$, $V$, $R$,
$I$, $J$, $H$, $K$, $L$ and the SLOAN/SDSS magnitudes in the AB
system, named $u_{sdss}$, $g_{sdss}$, $r_{sdss}$, $i_{sdss}$, and
$z_{sdss}$.

\begin{table*}
\begin{center}
\caption{Absolute Magnitudes evolution along the time/redshift for the grid of models in Johnson and SDSS/SLOAN system}
\begin{tabular}{rrrrrrrrrrr}
\hline
$nt$  & $dis$ &   $t$ & $z$ &  $Radius$  & $Area$ & $UV_{1}$ & $UV_{2}$ & $U$ & $B$ & $V$ \\ 
\hline
   4 &   28 &   13.201 &   0.00 &   0.0 &    19  &   -18.920 &   -18.115  &   -17.796 &   -18.015 &  -18.827\\  
   4 &   28 &   13.201 &   0.00 &   2.0 &    13  &   -18.151 &   -17.351  &   -17.130 &   -17.322 &  -18.112\\  
   4 &   28 &   13.201 &   0.00 &   4.0 &    25  &   -19.109 &   -18.301  &   -18.021 &   -18.228 &  -19.034\\  
   4 &   28 &   13.201 &   0.00 &   6.0 &    38  &   -19.953 &   -18.920  &   -18.602 &   -18.749 &  -19.491\\  
   4 &   28 &   13.201 &   0.00 &   8.0 &    50  &   -20.669 &   -19.358  &   -18.983 &   -19.016 &  -19.618\\  
   4 &   28 &   13.201 &   0.00 &  10.0 &    63  &   -20.368 &   -18.970  &   -18.578 &   -18.561 &  -19.092\\  
   4 &   28 &   13.201 &   0.00 &  12.0 &    75  &   -19.554 &   -18.117  &   -17.733 &   -17.691 &  -18.183\\  
   4 &   28 &   13.201 &   0.00 &  14.0 &    88  &   -18.561 &   -17.088  &   -16.727 &   -16.661 &  -17.112\\  
   4 &   28 &   13.201 &   0.00 &  16.0 &   100  &   -17.443 &   -15.922  &   -15.580 &   -15.495 &  -15.894\\  
   4 &   28 &   13.201 &   0.00 &  18.0 &   110  &   -16.149 &   -14.502  &   -14.091 &   -13.969 &  -14.354\\  
   4 &   28 &   13.201 &   0.00 &  20.0 &   139  &   -14.297 &   -12.508  &   -11.986 &   -11.827 &  -12.212\\  
   4 &   28 &   13.201 &   0.00 &  22.0 &   140  &   -11.665 &    -9.836  &    -9.223 &    -9.054 &   -9.435\\  
   4 &   28 &   13.201 &   0.00 &  24.0 &   150E &    -8.545 &    -6.786  &    -6.154 &    -5.971 &   -6.321\\  
\hline
\label{mag}
\end{tabular}
\end{center}
\end{table*}

\begin{table*}
\addtocounter{table}{-1}
\begin{center}
\caption{Cont. Absolute Magnitudes evolution along the time/redshift for the grid of models in Johnson and SDSS/SLOAN system}
\begin{tabular}{rrrrrrrrrrr}
\hline
$R$      &     $ I$     &      $ J$    &     $H$     &     $K$   &      $L$   &     $u_{sdss}$ & $g_{sdss}$ &   $r_{sdss}$  &   $i_{sdss}$ &   $z_{sdss}$ \\
\hline
19.476 &   -20.100  &   -20.726  &   -21.295 &  -21.526 &  -21.673&   -16.963 &  -18.537 &  -19.135 &  -19.486&   -19.720\\  
18.748 &   -19.353  &   -19.959  &   -20.516 &  -20.741 &  -20.885&   -16.292 &  -17.832 &  -18.412 &  -18.749&   -18.969\\  
19.679 &   -20.295  &   -20.911  &   -21.476 &  -21.704 &  -21.849&   -17.188 &  -18.747 &  -19.340 &  -19.686&   -19.914\\  
20.103 &   -20.700  &   -21.307  &   -21.867 &  -22.095 &  -22.241&   -17.776 &  -19.235 &  -19.775 &  -20.097&   -20.316\\  
20.155 &   -20.701  &   -21.277  &   -21.819 &  -22.041 &  -22.187&   -18.161 &  -19.429 &  -19.852 &  -20.117&   -20.310\\  
19.585 &   -20.096  &   -20.647  &   -21.174 &  -21.392 &  -21.536&   -17.757 &  -18.938 &  -19.296 &  -19.526&   -19.700\\  
18.650 &   -19.136  &   -19.664  &   -20.175 &  -20.386 &  -20.528&   -16.912 &  -18.050 &  -18.371 &  -18.577&   -18.734\\  
17.548 &   -18.005  &   -18.507  &   -19.000 &  -19.204 &  -19.342&   -15.901 &  -17.000 &  -17.278 &  -17.460&   -17.598\\  
16.294 &   -16.718  &   -17.186  &   -17.647 &  -17.836 &  -17.969&   -14.745 &  -15.807 &  -16.036 &  -16.188&   -16.304\\  
14.753 &   -15.151  &   -15.582  &   -15.990 &  -16.151 &  -16.270&   -13.254 &  -14.269 &  -14.498 &  -14.638&   -14.728\\  
12.623 &   -13.018  &   -13.442  &   -13.834 &  -13.985 &  -14.096&   -11.157 &  -12.122 &  -12.364 &  -12.511&   -12.592\\  
-9.872 &    10.303  &   -10.750  &   -11.184 &  -11.358 &  -11.478&    -8.399 &   -9.346 &   -9.600 &   -9.781&    -9.884\\  
-6.757 &    -7.202  &    -7.658  &    -8.119 &   -8.308 &   -8.435&    -5.333 &   -6.247 &   -6.482 &   -6.674&    -6.786\\  
\hline
\end{tabular}
\end{center}
\end{table*}

\begin{figure*}
\centering
\includegraphics[width=\textwidth,angle=0]{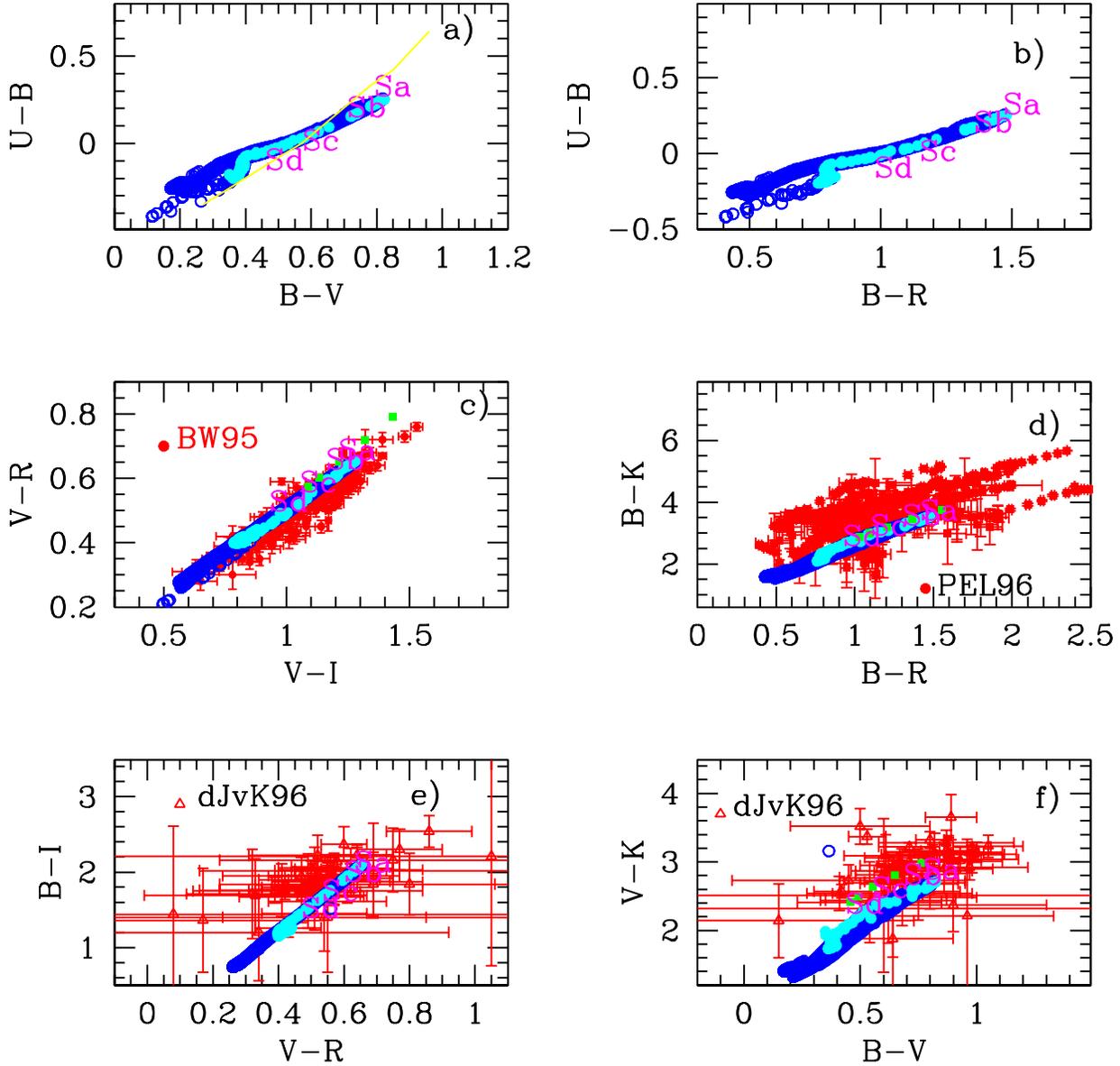} 
\caption{Color-color diagrams. Colors $U-B$ {\sl vs} $B-V$, $U-B$ {\sl
vs} $B-R$, $V-R$ {\sl vs} $V-I$, $B-K$ {\sl vs} $B-R$, $B-I$ {\sl vs}
$V-R$ and $V-K$ {\sl vs} $B-V$ are represented as blue dots and
compared with observational data.  The standard values for typical
galaxies along the Hubble Sequence as Sa, Sb, Sc, Sd are taken from
\citet{pog97} and shown as in magenta letters. In the panel $U-B$ {\sl
vs} $B-V$, the yellow line corresponds to data from \citet{vit96}, the
red dots are from \citet[][ BW95]{buta95} in $V-R$ {\sl vs} $V-I$,
from \citet[][ PEL96]{pel96} in panel $B-K$ {\sl vs} $B-R$, and from
\citet[][ dJvK96]{djong96} in $B-I$ {\sl vs} $V-R$ and $V-K$ {\sl vs}
$B-V$ panels.}
\label{color-color}
\end{figure*}

We may check that our results for the whole grid are reasonable
comparing them with color-color diagrams as it is shown in
Fig.~\ref{color-color}.  Colors $U-B$ {\sl vs} $B-V$, $U-B$ {\sl vs}
$B-R$, $V-R$ {\sl vs}$V-I$, $B-K$ {\sl vs} $B-R$, $B-I$ {\sl vs} $V-R$
and $V-K$ {\sl vs} $B-V$ are represented as blue dots for all regions
and galaxies modeled in this works and compared with observational
data from \citet{buta95} in panel $V-R$ {\sl vs} $V-I$, from
\citet{pel96} in panel $B-K$ {\sl vs} $B-R$, and from \citet{djong96}
in $B-I$ {\sl vs} $V-R$ and $V-K$ {\sl vs} $B-V$. The standard values
for typical galaxies along the Hubble Sequence as Sa, Sb, Sc, Sd taken
from \citet{pog97} are also shown, labeled in magenta.  In panel $U-B$
vs $B-V$ the yellow line corresponds to data from \citet{vit96}. Green
dots are from \citet{buz05}.  The cyan ones are the results
corresponding to the galaxies from Table~\ref{examples}.  In fact the
dispersion of data is quite large, in particular in the two bottom
panels.  Probably due to the contribution of the emission lines, which
move the points of the standard locus for galaxies in a orthogonal
way, such as we have demonstrated in \citet{mar12,gar13}, where we
added the contribution of the emission lines to the broad band colors
in single stellar populations and in star-forming galaxies models.

\subsection{Brightness and color radial profiles of disks}

As we known the luminosity of each radial region, we may compute the
surface brightness as $\rm mag\,arcsec^{-2}$. By assuming that our theoretical
galaxies are located at 10\,pc, the brightness is:
\begin{equation}
\mu=M+21.57+\log{Area}
\end{equation}
where $Area$ is the area of each radial region in $\rm pc^{2}$, and the
constant value 21.57 is  $2.5\times\log{\rm Fcon}$, where Fcon is the 
factor to convert  pc to arcsec.

As said before, we have not computed apparent magnitudes in each redshift
and then the relation luminosity distance-redshift is not necessary and
the redshift of the wavelength is not taken into account.

\begin{figure*}
\includegraphics[width=\textwidth,angle=0]{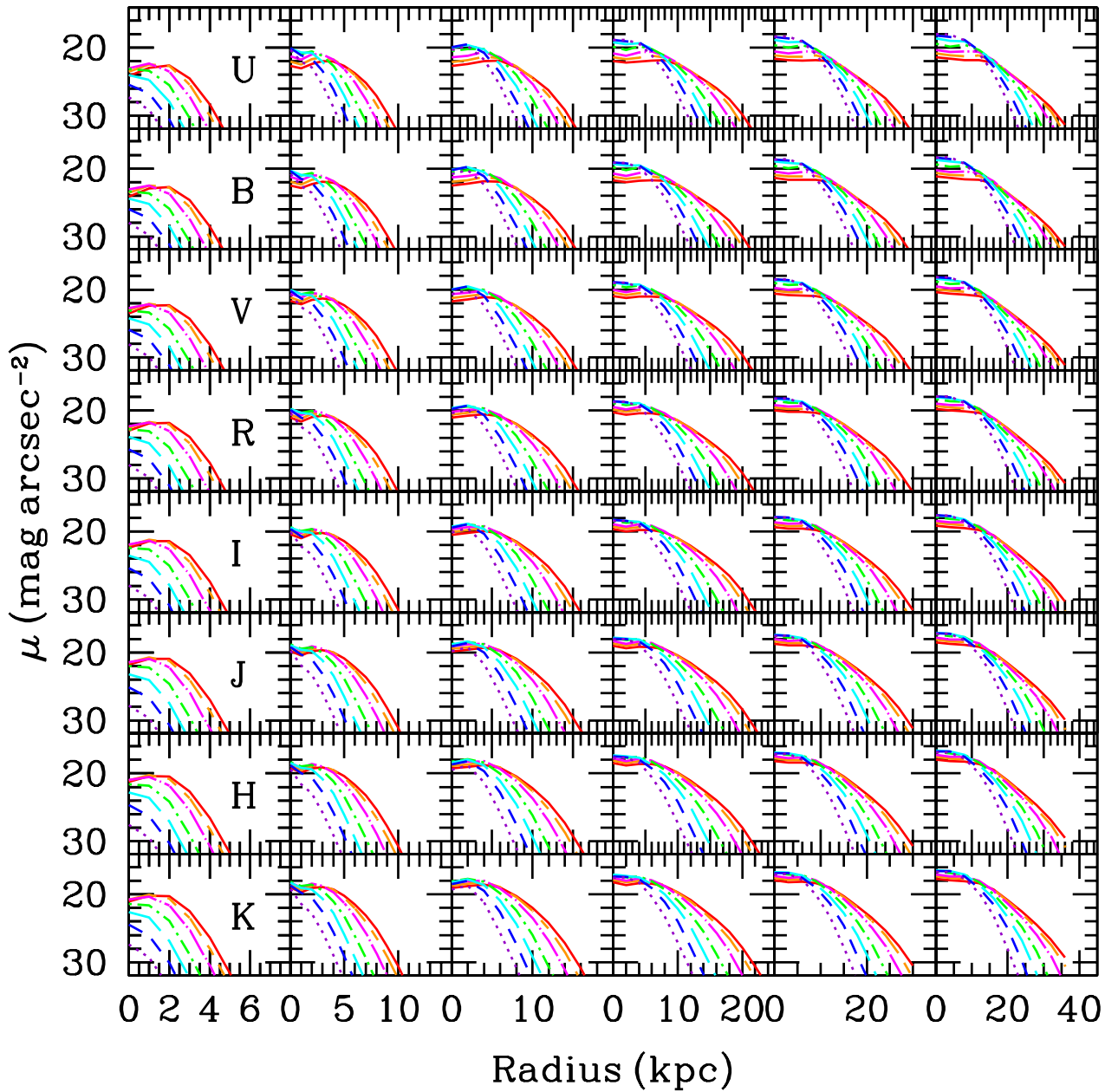} 
\caption{Evolution with redshift of surface brightness profiles in
$U$, $B$, $V$, $R$, $I$, $J$, $H$, and $K$ bands. Each row shows the
results for a theoretical galaxy of the Table~\ref{examples}, from the
galaxy, with $Vmax=78\,\rm km\,s^{-1}$, to the most massive on the
bottom for $Vmax=290\,\rm km\,s^{-1}$. Lines purple, blue, cyan,
green, magenta, orange and red are for $z=$5, 4, 3, 2, 1, 0.4 and 0,
respectively.}
\label{mag_z}
\end{figure*}

\begin{figure*}
\includegraphics[width=0.8\textwidth,angle=-90]{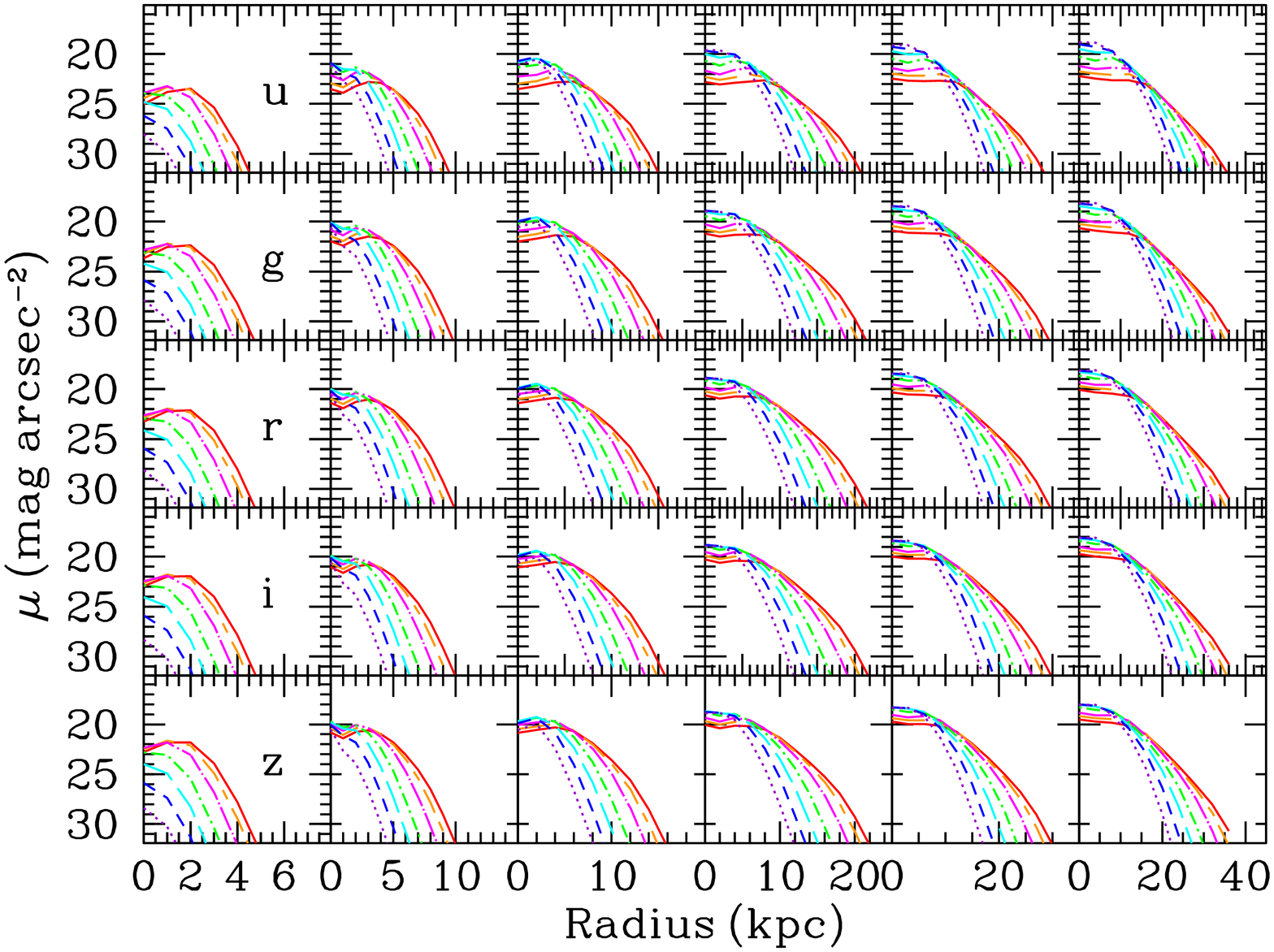} 
\caption{Evolution with redshift of surface brightness profiles for
SDSS/SLOAN magnitudes $u_{sdss}$, $g_{sdss}$, $r_{sdss}$, $i_{sdss}$
and $z_{sdss}$. Each row shows the results for a theoretical galaxy
of the Table~\ref{examples}, from the galaxy, with $Vmax=78\,\rm
km\,s^{-1}$, to the most massive on the bottom for $Vmax=290\,\rm
km\,s^{-1}$. Lines purple, blue,cyan,green,magenta, orange and red
are for $z=$ 5, 4, 3, 2, 1, 0.4 and 0, respectively.}
\label{mag_sloan_z}
\end{figure*}

Brightness radial profiles are shown in Fig.~\ref{mag_z} for the same
6 galaxies of the Table~\ref{examples} than Fig.6 and 8.  Each column
represents a galaxy, from the smallest one ($Vmax=78\,\rm km\,s^{-1}$)
at the left, to the most massive ($Vmax=290\,\rm km\,s^{-1}$) at the
right.  Brightness in bands $U$, $B$, $V$, $R$, $I$, $J$, $H$ and $K$
of the Johnson system are given from top to bottom. In
Fig.~\ref{mag_sloan_z} we show similar radial profiles brightness
radial profiles for bands $u_{sdss}$, $g_{sdss}$, $r_{sdss}$,
$i_{sdss}$, and $z_{sdss}$ in the SDSS/SLOAN system. In each panel the
same 7 redshifts than in previous figures 6 and 8 are shown with the
same color coding.  The profiles show the usual exponential shapes
except for the central/inner regions where a flattening is
evident. The results for the disks at $z=0$ are similar to the
observed ones.  The value of $\mu=21-22\rm\,mag\,arcsec^{-2}$ observed
as a common value in the center of most galaxies \citep{free70} is
found with our models.  Profiles are steeper for the highest
redshifts, being galaxies less luminous, and disks smaller. As the
galaxy evolves, more stellar mass appears and more extended in the
disk, doing the profiles more luminous and extended. Thus, in K-band,
the radius $R25$, defined as this one where $\mu=25\rm
\,mag\,arcsec^{-2}$, is in the most massive galaxy, $\sim 18$ \,kpc
for $ z=5$, and is $> 30$\, kpc for the present time. While for the
left column galaxy, $R25 \sim 1$ \,kpc for $z=3$ and $R25 \sim 4$
\,kpc for $z=0$ but the brightness do not reach this level at higher
redshifts than 3.

\begin{figure}
\centering
\includegraphics[width=0.35\textwidth,angle=-90]{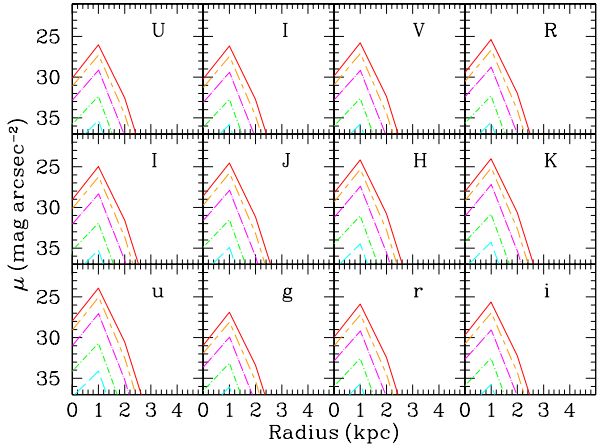} 
\caption{Evolution of surface brightness profiles for Johnson
($U$,$B$,$V$,$R$,$I$,$J$,$H$, and $K$) and SDSS/SLOAN ($u_{sdss}$,
$g_{sdss}$, $r_{sdss}$, and $i_{sdss}$) magnitudes radial
distributions for the least massive galaxy of the Table~\ref{examples}
with $Vmax=48\,\rm km\,s^{-1}$.}
\label{mag_lowmass}
\end{figure}

Again we show separately in Fig.~\ref{mag_lowmass} the surface
brightness profiles for the lowest mass galaxy of our
Table~\ref{examples}: It has a very low luminosity in all bands and
surface brightness that are difficult to be observed even in the local
Universe. Only the region around ~1\,kpc of distance might be observed
for $z< 2$.

\begin{figure*}
\centering
\includegraphics[width=0.7\textwidth,angle=-90]{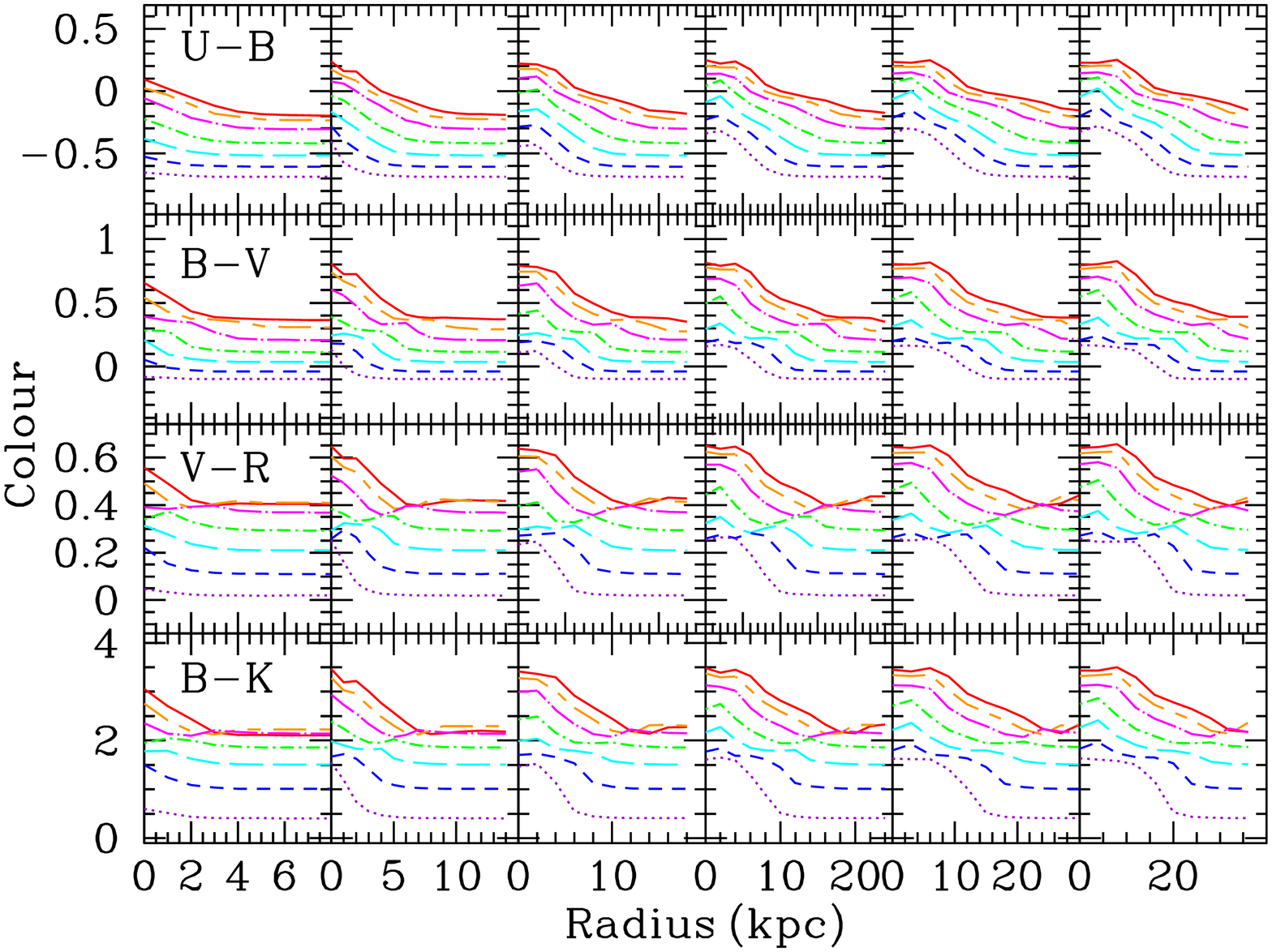} 
\caption{Evolution of colors radial profiles for $U-B$, $B-V$, $V-R$
and $R-I$. As in previous figures each color represents a redshift
while each column is a different theoretical galaxy of the
Table~\ref{examples}, from the galaxy, with $Vmax=78\,\rm km\,s^{-1}$,
to the most massive on the bottom for $Vmax=290\,\rm km\,s^{-1}$.}
\label{color_J}
\end{figure*}

\begin{figure*}
\centering
\includegraphics[width=0.7\textwidth,angle=-90]{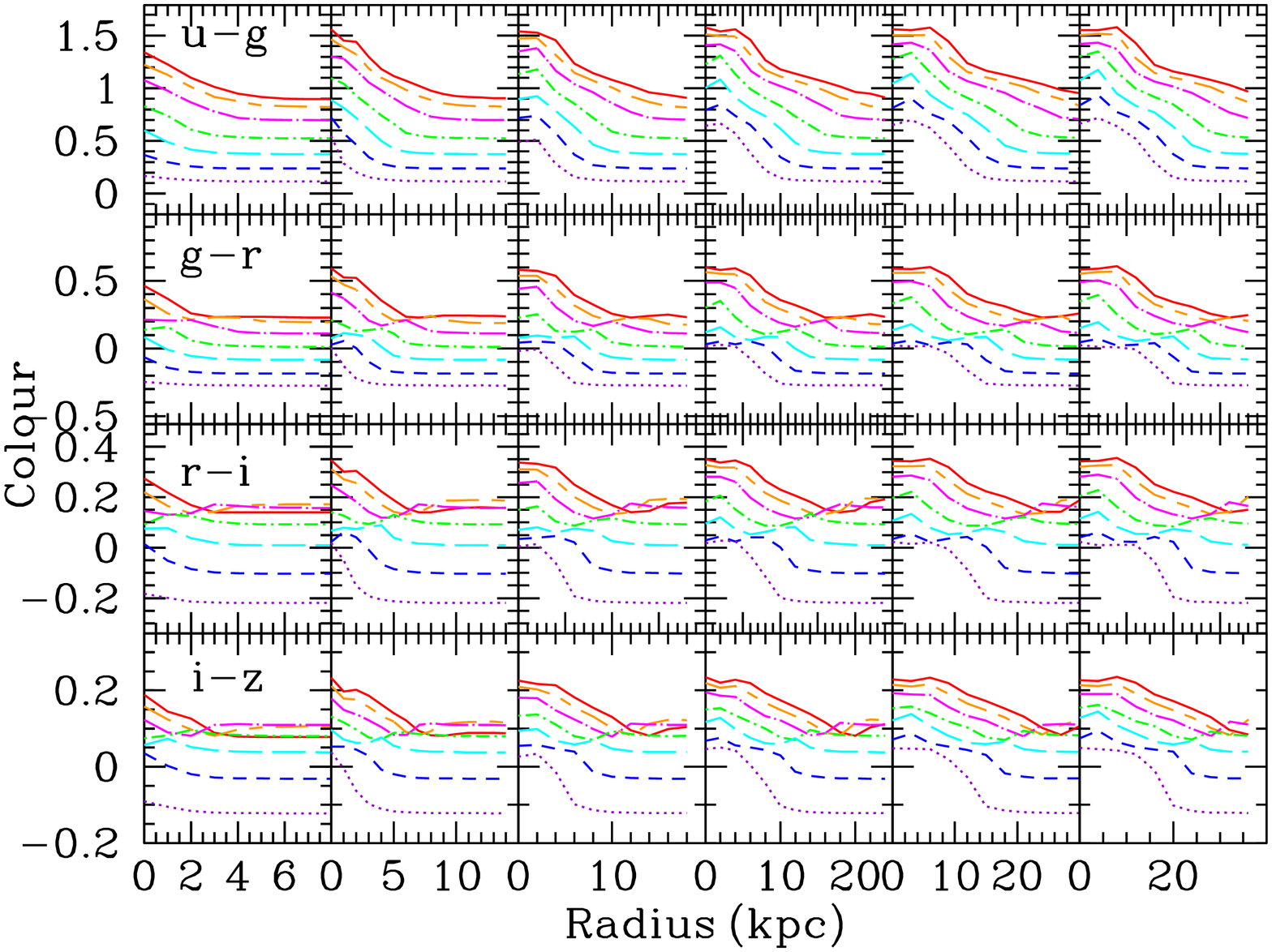} 
\caption{Evolution of colors radial profiles for
$u_{sdds}-g_{sdds}$,$g_{sdds}-r_{sdds}$, $r_{sdds}-i_{sdds}$,
$i_{sdds}-z_{sdds}$.  As in previous figures each color represents a
redshift while each column is a different theoretical galaxy of the
Table~\ref{examples}, from the galaxy, with $Vmax=78\,\rm km\,s^{-1}$,
to the most massive on the bottom for $Vmax=290\,\rm km\,s^{-1}$.}
\label{color_S}
\end{figure*}

The radial profiles for some colors are shown in Fig.~\ref{color_J}
and \ref{color_S} with similar columns for the same 6 galaxies from
Table~{examples} as before, the lowest mass at left, the most massive
one to the right.  These radial distributions do not show uniform
evolution doing evident that not all bands evolve equally. Thus,
colors $V-R$, $B-K$ in Fig.~\ref{color_J} or $r_{sdss}-i_{sdss}$ and
$i_{sdss}-z_{sdss}$ in Fig.~\ref{color_S} show at $z=2-3$ an increase
in some place along the disk.

\section{Discussion}

\begin{figure}
\centering
\includegraphics[width=0.5\textwidth,angle=0]{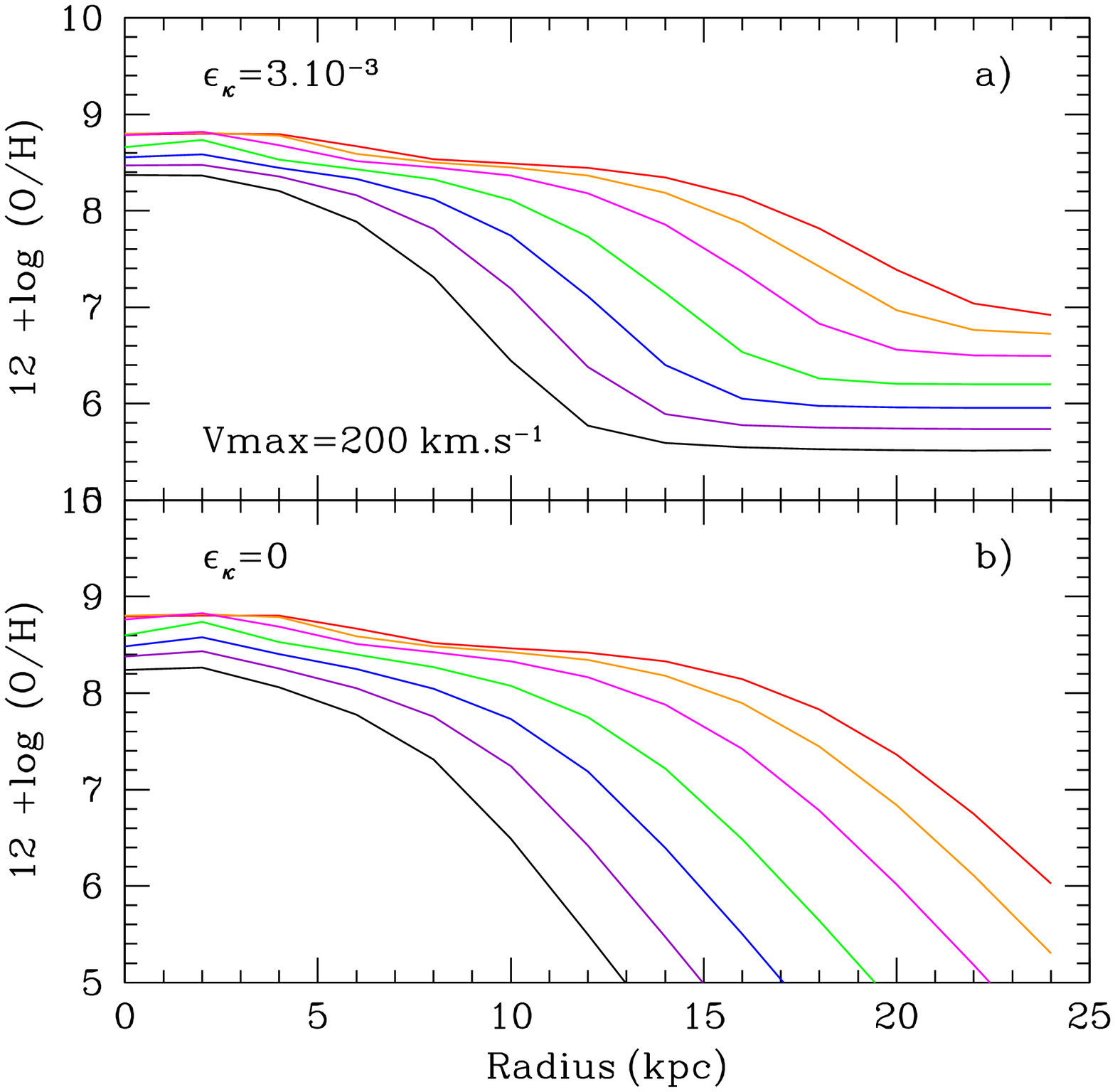} 
\caption{Evolution of the radial distributions of oxygen abundance for
a MWG-like galaxy (number 5 from Table\ref{examples}.  a) With a star
formation efficiency in the halo, $\epsilon_{\kappa}=0.003$. b). Without
star formation in the halo $\epsilon_{\kappa}=0$.}
\label{ohr}
\end{figure}

One of the things which arises from these models is that elemental
abundances do not show exponential radial distributions and therefore
it is not easy to fit a straight line in the logarithmic scale and to
obtain a radial gradient. The distributions are always flatter on the
inner regions than in the disk. It is in the disk region between the
bulge and the optical radius (around 2 times the effective radius)
where a radial gradient may be actually be well defined. In the outer
regions of the disk the distributions are flatter again, which is in
agreement with recent observations from the CALIFA survey
\citep{san13}.  This flattening is more evident, at shorter radii, for
early times or highest redshifts.  By taking into account that at
these same time/redshifts the stellar profiles continue being steep,
it is evident that the abundances do not proceed from the stellar
production in the disk. Since the radial gradient is created by the
ratio $\Psi/f$, one possibility is that the infall causes en
enrichment in the outer parts of the disks. We must remain that in our
models the halos create stars too, with an efficiency
$\epsilon_{\kappa}=0.003$.  With this value it is possible reproduce the
star formation history and the age-metallicity relation of the
halo. Since the infall rate is very high in the inner regions of the
disk, the star formation in the inner halo occurs during a very short
time, then the gas falls towards the disk, and the star formation
in the halo stops. Thus, stars in the halo will be old and with low
metallicities, as observed. On the contrary the infall of gas takes
places during a longer time in the outer disk, and therefore the
corresponding halo maintains a quantity of gas which allows to have a
star formation rate at a certain level along the galaxy life. Thus the
gas infalling in the disk may be enriched, in a relative term,
compared with the extremely low abundances of the outer disk.  In
order to check this possibility we have computed a model for a
theoretical galaxy similar to MWG, which would be as the number 5 in
Table\ref{examples}, without star formation rate, that is with
$\epsilon_{\kappa}=0$. Resulting radial distributions for both cases may be
compared in Fig.~\ref{ohr}. In a) we show the standard results, already
shown in Fig.8 with the other galaxy examples. In b) we have the
same model with $\epsilon_{\kappa}=0$. In the first case the flattening of
the radial distributions for the outer disk is evident for all
redshifts, although more clear in the highest ones, while in the
bottom panel no flattening in this region appears. The flattening in
the inner disk is similar in both cases.

\begin{figure}
\centering
\includegraphics[width=0.35\textwidth,angle=-90]{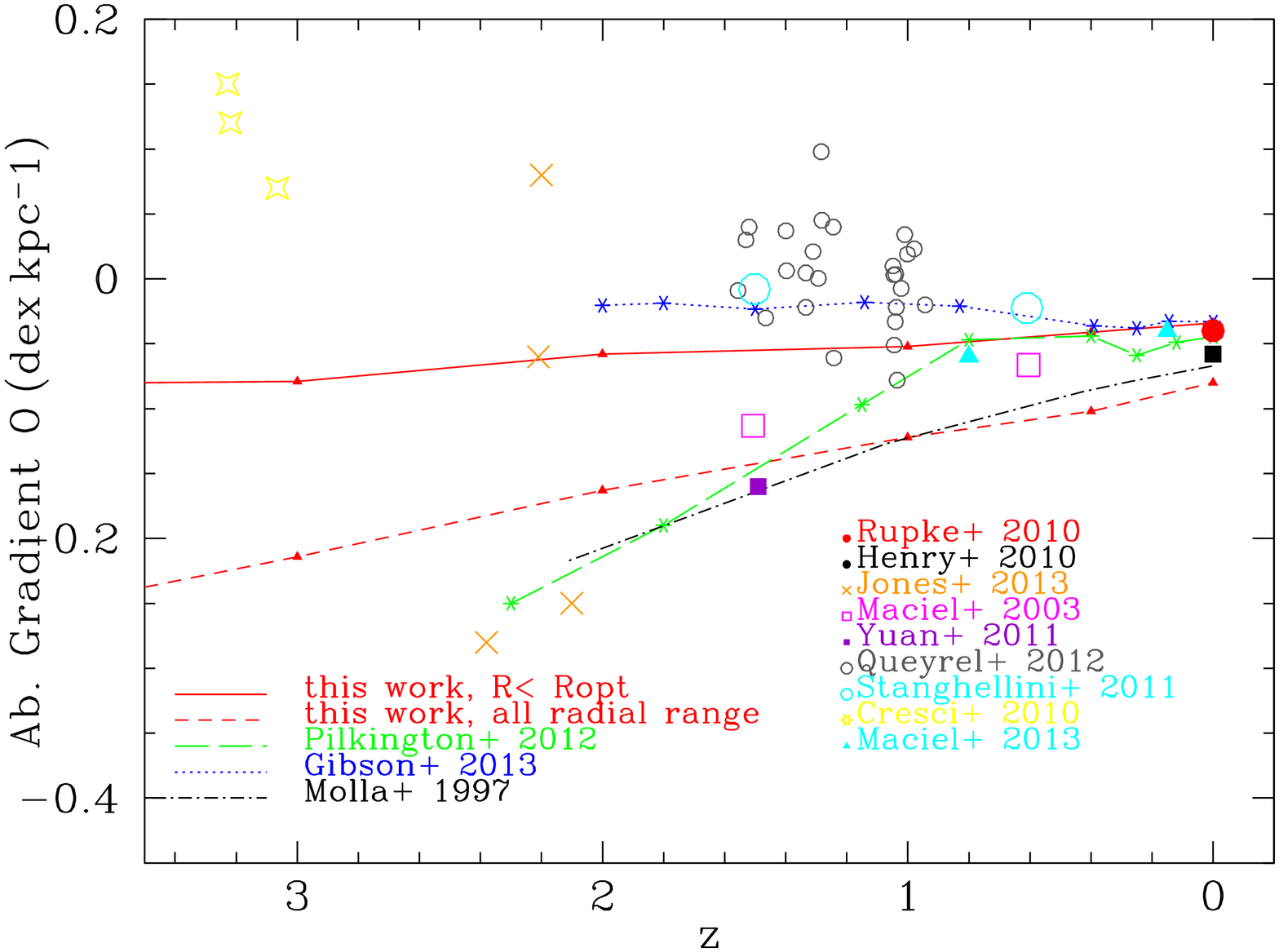} 
\caption{Evolution of the radial gradient of oxygen abundance for a
  MWG-like galaxy (number 5 from Table\ref{examples} along the
  redshift compared with observations. Dots are data are from
  \citet{mac03,rupke10,stan10,cres10,yuan11,quey12,jon13,mac13} as
  labeled, while lines are simulations from \citet{pil12, gib13}, and our
  old model results for MWG \citep{mol97}, as given in the plot. For
  this work we have two lines, one obtained by using all the radial
  range and other only for disk within the optical radius, that are
  shown by the short-dashed and solid lines, respectively.}
\label{grad}
\end{figure}

The second result is that radial gradient flattens with the evolution
for all galaxies. However the rate with which this occurs is not the
same for all of them. Massive galaxies evolve more rapidly flattening
very quickly their radial distributions of abundances.  Low mass
galaxies on the contrary maintain a steep radial gradient for a longer
time. On the other hand, this is a generic result when all radial
range of the disk is used to compute the radial gradient. By taking
into account that the distributions of abundances have not an unique
slope, as we have explained above, we might to select different ranges
to compute this gradient.  This if we calculate the gradient for all
the spatial range, we obtain the red dashed line in Fig~\ref{grad} for
a MWG-like model, similar to our results in \citet{mol97}.  The radial
gradient decreases with the evolution. If we select a restricted
radial range, computing it only for $Radius<Ropt$ ( which increases
with redshift decreasing), then we obtain the solid line results which
show a smaller absolute value and less evolution along redshift.
It is necessary to remind that other galaxies will have their own
radial gradient evolution since each galaxy may evolve in a different
way.
 
We may compare these results with observational data which are now
being published.  We do that in Fig.~\ref{grad} where data from
\citet{mac03,rupke10,stan10,cres10,yuan11,quey12,jon13,mac13} are
shown. It is evident that not all of them give the same result. Some
authors claim that the radial gradient are steeper for early
evolutionary times, while others found flat or even positive radial
gradient, and try to interpret these results with generic scenarios
about the formation of galaxies. It is necessary to say again that not
all galaxies evolves in the same way, and, furthermore, not all the
observations measure the same thing. The radial range of the
measurements is important as we have shown before. There are other
important observational effects, such as the angular resolution, the
signal to noise, or the annular binning that may change the obtained
radial gradient, such as it is demonstrated in \citet{yuan13}. It is
necessary to take care of how using these high redshift observations
before to extract conclusions.  Maybe it is a more sure method
to study the planetary nebula that give to us the radial gradient that
a galaxy had some time ago, such as \citet{mac03,mac13} and their
group do.

\section{Conclusions}
 
We have shown a complete grid of chemo-spectro-photometric evolution
models calculated for spiral and irregular galaxies. The evolution
with redshift is given in the rest-frame of galaxies. We obtain the
evolution of the radial gradient of abundances with higher abundances
in the inner regions of disks than in the outer ones. These radial
gradients flatten with decreasing redshifts, but always there are some
outer regions that show no radial gradient, or it is flatter than in
the inner disk. These regions are located more far than the center 
when the evolution takes place. We have also presented the photometric
evolution for this same set of theoretical galaxies, given the surface
brightness profiles at different evolutionary times or redshifts.
Using the surface density of atomic, molecular or stellar masses, and
these surface brightness profiles we may predict observational limits
for these quantities for different redshifts. We may also check that 
the flat radial gradients  of abundances in the outer disks do not
correspond to a similar flattening of the surface brightness profiles, and 
therefore the abundances in these regions do not arise  by the stellar
populations in the disk. We suggest that they are the result of the
infall of gas coming from a halo who is in that moment more enriched
than the one in the disk.  This indicates that the infall law of gas
which forms-out the disk has important consequences in the predicted
observational characteristics of galaxies at high redshift. Therefore
to analyze other possible infall laws is essential and we will do that in a next future. 

\section{Acknowledgments}
This work has been supported by DGICYT grant AYA2010-21887-C04-02 and
04. Also, partial support from the Comunidad de Madrid under grant CAM
S2009/ESP-1496 (AstroMadrid) is grateful. M.Moll\'{a} thanks the kind
hospitality and wonderful welcome of the Instituto de Astronomia,
Geof\'{\i}sica e Ci\^{e}ncias Atmosf\'{e}ricas in Sao Paulo, where this
work has been finished. We thank an anonymous referee for many useful comments and suggestions
that have greatly improved this paper.

\end{document}